\documentclass[preprint]{aastex63}

\usepackage{verbatim}
\usepackage{cprotect}
\received{\today}
\revised{\today}
\accepted{\today}
\submitjournal{AJ}

\shortauthors{Quarles et al.}

\begin{document}

\title{Exomoons in Systems with a Strong Perturber: Applications to $\alpha$ Cen AB}

\correspondingauthor{Billy Quarles}
\email{billylquarles@gmail.com}

\author[0000-0002-9644-8330]{Billy Quarles}
\affil{Center for Relativistic Astrophysics, School of Physics, 
Georgia Institute of Technology, Atlanta, GA 30332 USA}

\author[0000-0002-1398-6302]{Siegfried Eggl}
\affil{Department of Astronomy, University of Washington, Seattle, WA 98195, USA}
\affil{Department of Aerospace Engineering, University of Illinois at Urbana-Champaign, Urbana, IL 61801, USA}
\affil{IMCCE, Observatoire de Paris, Paris, France}

\author[0000-0003-0216-559X]{Marialis Rosario-Franco}
\affil{National Radio Astronomy Observatory, Socorro, New Mexico, USA}
\affil{University of Texas at Arlington, Department of Physics, Arlington, Texas, USA}

\author[0000-0001-8308-0808]{Gongjie Li}
\affil{Center for Relativistic Astrophysics, School of Physics, 
Georgia Institute of Technology, Atlanta, GA 30332 USA}



\begin{abstract}
{The presence of a stellar companion can place constraints on occurrence and orbital evolution of satellites orbiting exoplanets, i.e., exomoons. In this work we revise earlier orbital stability limits for retrograde orbits in the case of a three body system consisting of star-planet-satellite. {The revised limit} reads $a_{\rm sat}^{\rm crit} \approx 0.668(1-1.236e_{\rm p})$ for $e_p \leq 0.8$ in units of the Hill Radius and represents the lower critical orbit as a function of the planetary eccentricity $e_{\rm p}$.  A similar formula is determined for exomoons hosted by planets in binary star systems, where $e_{\rm p}$ is replaced with the components of free and forced eccentricity from secular orbit evolution theory. By exploring the dynamics of putative exomoons in $\alpha$ Centauri AB we find that the outer stability limit can be much less than half the Hill Radius due to oscillations in the planetary orbital eccentricity caused by the gravitational interaction with the binary star.} We show, furthermore, how the resulting truncation of the outer stability limit can affect the outward tidal migration and potential observability of exomoons through transit timing variations (TTVs). Typical TTV {(RMS)} amplitudes induced by exomoons in binary systems are $\lesssim$10 min and appear more likely for planets orbiting the less massive stellar component.

\end{abstract}



\section{Introduction}

The detection of the first exoplanets \citep{Wolszczan1992,Mayor1995} established unequivocally that planets like those in the solar system can form around other stars and be detected using existing technology.  However, there are more moons (satellites) within the solar system than planets and if planets are abundant, then moons should be as well.  The detection of exomoons is the next frontier.  After the first planets, the method of transits was identified as a viable detection method for them \citep{Sartoretti1999}.  Later, \cite{Holman2005} and \cite{Agol2005} showed the detection of Earth-like planets was possible using transit-timing methods, which implied that the era of exomoon detection could also be forthcoming \citep{Cabrera2007}.  

\cite{Kipping2009a,Kipping2009b} showed the architectures of planet-moon pairs that would produce the strongest observational signatures through transit timing and duration variations.  {Recently, \cite{Kipping2020} and \cite{Kipping2021} detailed the expected limitations on observational signatures using transit timing variations.  Finding such signatures would theoretically be possible through observations with the {\it Kepler} Space Telescope,} but an extensive analysis of its data did not readily confirm any exomoons \citep{Kipping2012,Kipping2013a,Kipping2013b,Kipping2014,Kipping2015b}.  Careful analysis did rule out several false positives, including one candidate for Kepler-90g \citep{Kipping2015a}.  Alternative methods from transit timing were also proposed for the detection of exomoons, including using the averaged light curves \citep{Simon2012}, optimizing with respect to the orbital sampling effect \citep{Heller2014,Heller2016,Hippke2015}, Doppler monitoring of directly imaged exoplanets \citep{Agol2015,Vanderburg2018}, or examining the radio emissions from giant exoplanets \citep{Noyola2014,Noyola2016}.

The stability of exomoons is also a major concern when prescribing the various means of detection.  Although largely unconstrained, tidal interactions play a significant role in determining the lifetimes of solar system moons and those orbiting exoplanets \citep{Barnes2002,lainey2020resonance}.  However, \cite{Sasaki2012,Sasaki2014} placed constraints on the long-term evolution of exomoons by exploring a wide range of tidal parameters (e.g,. tidal Love number and tidal quality factors).  Moreover, \cite{Namouni2010} and {\cite{Trani2020}} tracked the fates of exomoons that could have formed prior to substantial inward migration of their host planets and found the shrinking planetary Hill radius $R_{\rm H}$ to cause such moons to become unstable.  \cite{Spalding2016} expanded this analysis to identify a mechanism within the evection resonance that causes significant eccentricity growth, where collisions can occur with the host planet or eventual tidal breakup due to crossings inside the Roche limit.  Eccentricity growth can also be induced through interactions with neighboring planets, where the strength of the interactions can be scaled by the host planet's Hill radius \citep{Payne2013}.  Furthermore, Lidov-Kozai oscillations with a tilted exomoon orbit can promote eccentricity growth of the host planet, where the precession rate from short range forces must greatly exceed the precession from the Lidov-Kozai interaction to ensure exomoon stability \citep{Grishin2018}.     \cite{Sucerquia2019} showed that exomoons can be captured by the host star as a `ploonet', where the moons can escape instead of collapsing onto their host planet during migration and can complicate the detection of exoplanets.  \cite{Domingos2006} produced estimates for the outer stability boundary for prograde and retrograde exomoons based upon numerical simulations and scaled their results relative to the planetary Hill radius $R_{\rm H}$, but those results overestimate the outer stability boundary by $\sim$20\% for prograde orbits \citep{Rosario-Franco2020} and represent the upper boundary of a transition region for stability \citep{Dvorak1986}.

Exoplanet searches have also uncovered planets that orbit two stars in either a {\it circumstellar} (e.g., $\gamma$ Cep \citep{Campbell1988,Hatzes2003} or {\it circumbinary} configuration (i.e., Kepler-16 \citep{Doyle2011}, Kepler-34 \& 35 \citep{Welsh2012}, Kepler-38 \cite{Orosz2012b}, Kepler-47 \citep{Orosz2012a,Orosz2019}, Kepler-64 \citep{Schwamb2013}, Kepler-413 \citep{Kostov2013}, Kepler-453 \citep{Welsh2012}, Kepler-1647 \citep{Kostov2016}, Kepler-1661 \citep{Socia2020}, TOI-1338 \citep{Kostov2020}).  Each of these systems hosts a giant planet. In terms of its bulk composition (i.e., planetary mass and radius) those planets range from Neptune-like to Jupiter-like.  \cite{Quarles2012} and \cite{Hamers2018a} showed the potential for the circumbinary giant planets to host moons.  However, an equivalent {study} has not been performed for circumstellar systems, which is the focus of this work.

The formation of exomoons around circumstellar planets in binaries could proceed through similar processes put forward for the solar system \citep[i.e.,][]{Canup2006} or potentially through tidal capture as has been proposed for the exomoon candidate Kepler-1625b-I \citep{Hamers2018b}.  The stability of the circumstellar host planets can also be influenced {by} the secular forcing from the secondary star \citep{andrade2017secular} or mean motion resonances that can eject the planet from the system \citep{Quarles2020}.  Due to the formation scenarios and constraints for orbital stability, the exomoons orbiting these giant planets can do so in a prograde or retrograde direction relative to the orbital direction of the host planet.  Jupiter has captured many moons and their retrograde orbits are taken as a tell-tale sign of this evolution \citep{Jewitt2007}. 

In this work, we apply an analytic framework for the stability of exomoons that orbit planets in binary star systems using the secular forced eccentricity and overlap of mean motion resonances in Section \ref{sec:analytic}. In section \ref{sec:numerical} we present the results of N-body simulations of a hypothetical planet-moon system in $\alpha$ Centauri AB and compare the numerical result to the predictions of the theoretical framework.  In addition, we test whether the outward tidal migration is affected by the stellar companion in Section \ref{sec:tides}.  Section \ref{sec:observ} explores the observational signature of exomoons through transit timing variations for exomoons in binary systems and estimates the number of potential systems using the statistics for the stellar binary population.  Section \ref{sec:conclusions} summarizes the key findings of our work.

\section{Analytic Criteria for Exomoon Stability with a Strong Perturber}\label{sec:analytic}
\subsection{Secular Forced Eccentricity}
The secular forced eccentricity from the three body problem can provide a simple analytical estimate on the region of stability around a planet orbiting one star of a binary star system. In such a system, the exomoon exists in a three body hierarchy, where it orbits a planet that in turn orbits its host star. In addition, a secondary star orbits the center of mass of three body system at such a distance so that the planetary orbit remains stable to produce a hierarchical four body configuration. 

Various aspects of the stability of hierarchical four body configurations have been investigated in some detail \citep[e.g.][and references therein.]{milani1983stability,liu2019stability}. Most approaches are based on Jacobi (quasi) integrals, $c^2 E$, where $c$ is the angular momentum and $E$ the total energy of the system \citep[i.e.,][]{Marchal1982}. Jacobi integrals are constants of motion in the circular restricted three body problem related to zero velocity curves. In order for a three body configuration to be Hill stable, the value of $c^2 E$ must be less
than the critical value at the Lagrange point L1. If that is the case, the zero velocity surface prevents a (massless) exomoon from escaping its orbit around the planet. 
Although this concept can be shown to yield rigorous and sufficient criteria, derived expressions for stability limits are often lengthy and cumbersome to use, especially in the general three and four body problems where the Jacobi integrals are not constant.

We consider the hierarchical four body problem as two coupled three body problems: 1) planet-moon-host star and 2) planet-moon barycenter-host star-stellar companion.  In the first three body hierarchy, the planetary semimajor axis $a_{\rm p}$ relative to the host star is much larger than the moon's semimajor axis $a_{\rm sat}$ relative to the host planet (i.e., $a_{\rm p} \gg a_{\rm sat}$). Then, the radius of the Hill sphere around the planet depends on the planetary semimajor axis and masses through,
\begin{equation}
    R_{\rm H} = a_{\rm p} \left(\frac{M_{\rm p}+M_{\rm sat}}{3M_{\star}}\right)^{1/3},
	\label{eq:rH}
\end{equation}
where $M_{\rm p}$, $M_{\rm sat}$, and $M_{\star}$ are the masses of the host planet, moon and host star, respectively.  The above definition corresponds to the Hill radius for a circular planetary orbit. In a more general formula $a_{\rm p}$ is replaced by $r_{\rm p}$ which also includes the planetary eccentricity $e_{\rm p}$. As a consequence, $R_{\rm H}$ depends on the position of the planet on its orbit with a minimum at 
\begin{equation}
    R_{\rm H}^\prime = (1-e_{\rm p})R_{\rm H}.
\end{equation} Since the planet, moon and host star form a hierarchical triple, the semimajor axis of the moon around the planet remains approximately constant over time. To guarantee Hill stability of a moon around the planet, we have to identify pathways for the moon to escape the Hill sphere around the planet, which typically occurs when the planet is at periastron.

Under an external perturbation, the Hill radius $R_{\rm H}^\prime$ is modified as the planetary eccentricity evolves \citep{georgakarakos2003eccentricity}, where a simple model for the planetary orbital eccentricity evolution in a hierarchical system dates back to \citet{heppenheimer1978ecc}. In essence, the eccentricity of the planetary orbit can be decomposed into forced ($\epsilon$) and free ($\eta$) components \citep{andrade2017secular}. The former is determined by system parameters, such as the ratio of semimajor axes and the eccentricity of the stellar binary, and can be derived from initial conditions. 
In their simplest form the equations for the forced and free amplitudes of the eccentricity vector read as:
\begin{equation}
    \epsilon_{\rm p}\approx\frac{5}{4}\frac{a_{\rm p}}{a_{\rm bin}}\frac{e_{\rm bin}}{1-e_{\rm bin}^2},
	\label{eq:eforced}
\end{equation}
and
\begin{equation}
    \eta_{\rm p}\approx |e_{\rm p,o}-\epsilon_{\rm p}|,
	\label{eq:efree}
\end{equation}
where $e_{\rm p,o}$ and $a_{\rm p}$ are the initial eccentricity and semimajor axis of the planet ${\rm p}$. The semimajor axis and eccentricity of the stellar binary are denoted by $a_{\rm bin}$ and $e_{\rm bin}$, respectively.
The maximum planetary eccentricity ($e_{\rm p,max}$) can be estimated by adding the amplitudes of the forced and free components of the respective eccentricity vector,
\begin{equation}
    e_{\rm p,max}\approx \eta_{\rm p} + \epsilon_{\rm p}.
	\label{eq:emax}
\end{equation}

In our four body configuration, the influence of the secondary star induces variations in the planetary eccentricity as the the system evolves.  As a result, we determine the smallest extent of the planet's Hill radius as 
\begin{equation}
    R^\prime_{\rm H, min} =  a_{\rm p} (1-e_{\rm p,max}) \left(\frac{M_{\rm p}+M_{\rm sat}}{3M_{\star}}\right)^{1/3}.
	\label{eq:rHmin}
\end{equation}

To become unbound, a moon's orbital energy $h$ must exceed the gravitational potential $U$, where this can occur when $r_{\rm sat} \geq R^\prime_{H,min}/2$.  We ignore the moon's eccentricity $e_{\rm sat}$ because the amplitude of its forced eccentricity is small, $\epsilon_{\rm sat} \propto a_{\rm sat}/a_{\rm p} \ll 1$.  Thus, we can construct a simple stability criterion for moons around planets that orbit one star of a binary star system, namely 
\begin{equation}
    a_{\rm sat}^{\rm crit} = \frac{1}{2}R^\prime_{H,min} \approx \frac{1}{2} \left(1-\epsilon_{\rm p}-\eta_{\rm p} \right) R_{\rm H},
	\label{eq:stab}
\end{equation}
where $\epsilon_{\rm p}$ and $\eta_{\rm p}$ are the forced and free eccentricities of the planet (see equations \ref{eq:eforced} and \ref{eq:efree}), and $R_{\rm H}$ is the Hill radius for a circular planetary orbit.  Moons on stable orbits would have semimajor axes $a_{\rm sat} < a_{\rm sat}^{\rm crit}$. Note that in Equation \ref{eq:stab} $a_{\rm sat}^{\rm crit}$ is given in units of distance. For the remained of this article we will measure $a_{\rm sat}^{\rm crit}$ in units of $R_H$.

\subsection{Stability Limitations from Mean Motion Resonances}
For moderate planetary eccentricity ($e_p \gtrsim 0.3$) N:1 mean motion resonances (MMRs) between the planetary and satellite orbits add an additional component to the eccentricity evolution of the satellite.  For example, the 8:1 MMR occurs for $a_{\rm sat}\sim0.36$ R$_{\rm H}$ and the 9:1 MMR occurs for $a_{\rm sat}\sim0.\bar{3}$ R$_{\rm H}$ following $a_{\rm sat}^{N} = (3/N^2)^{1/3}$ R$_{\rm H}$ {\citep{Kipping2009a,Rosario-Franco2020}}.  The libration width of MMRs grows with the perturber's eccentricity (i.e., $e_p$), which can allow for the overlap of MMRs to destabilize a significant range in $a_{\rm sat}$ \citep{Murray1999,Mudryk2006,Morais2012}.  From \cite{Mardling2013}, we calculate the libration width $\Delta\sigma_N$of the N:1 MMRs using the following:

 \begin{equation} \label{eq:MMR}
     \Delta\sigma_N = \frac{6\mathcal{H}_{22}^{1/2}}{(2\pi)^{1/4}}
     \sqrt{\frac{M_\star}{M_T}+N^{2/3}\frac{M_p M_{sat}}{M_T^{2/3}(M_p+M_{sat})^{4/3}}}
     \frac{\sqrt{e_{sat}}}{e_p}\sqrt{1-\frac{13}{24}e_{sat}^2}(1-e_p^2)^{3/8}N^{3/4}\mathrm{e}^{-N\xi(e_p)/2},
 \end{equation}
where $\mathcal{H}_{22}=0.71$ is a scale factor from the spherical harmonic expansion of a pendulum model for resonance, $M_\star$ is the mass of the host star, $M_T$ is the total mass of the star-planet-satellite system, and $\xi(e_p) = \mathrm{cosh}^{-1}(1/e_p) - \sqrt{1-e_p^2}$.  Note that $\Delta\sigma_N=0$ for $e_{sat}=0$, which implies that circular satellite orbits are always stable in this model, and this is not the case, as noted by \cite{Mardling2013}.  When calculating $\Delta\sigma_N$ for circular orbits, we use $e_{\rm sat}=10^{-3}$ or $e_{\rm p} = 10^{-3}$ to avoid the complication {noted by \citeauthor{Mardling2013} or the singularity for $e_{\rm p}\rightarrow 0$, respectively.}

\begin{figure}
	\includegraphics[width=0.9\columnwidth]{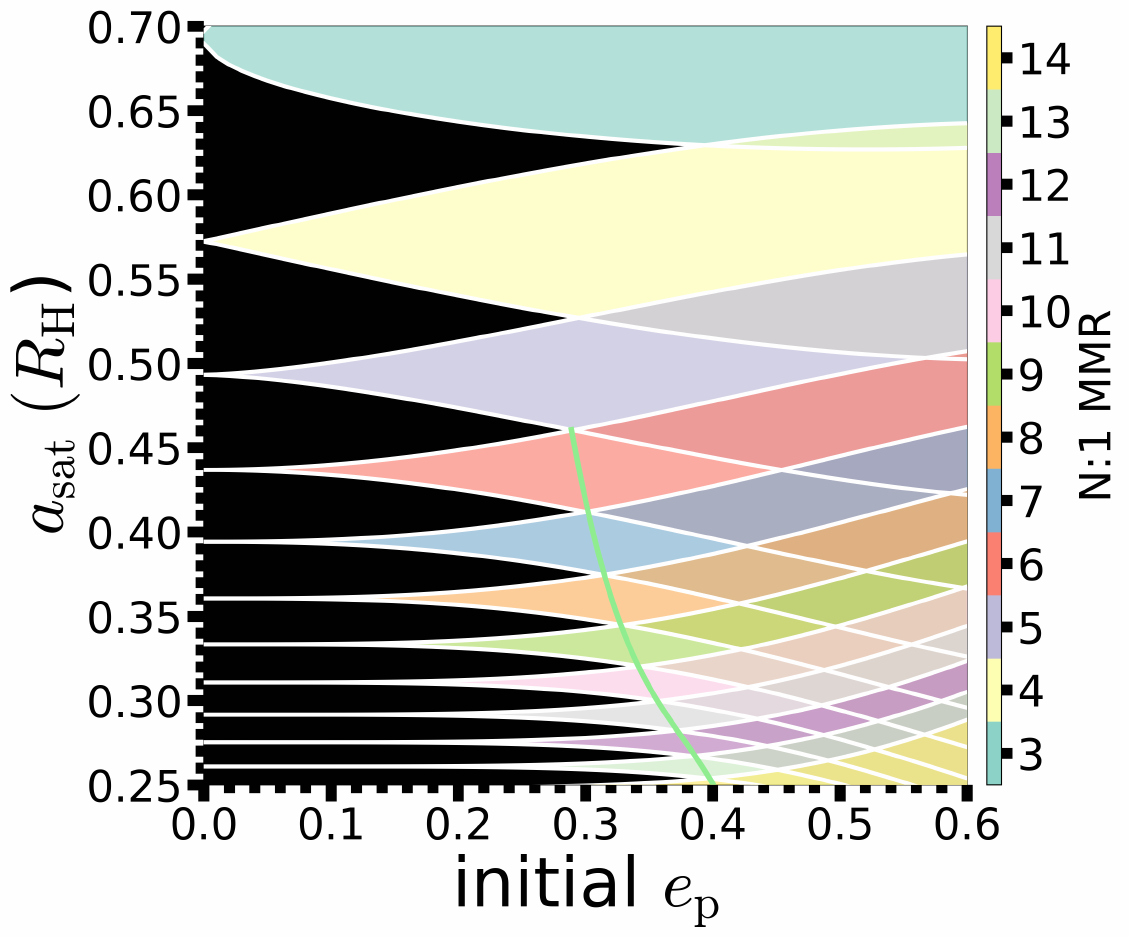}
    \caption{Map of the N:1 mean motion resonances (MMRs), where the color-code denotes the value of N for each MMR.  The regions between MMRs are colored black and could allow for stable satellite orbits under optimal conditions.   The light green curve connects the first point of intersection between adjacent MMRs and marks a stability boundary within the three body problem.}
    \label{fig:moon_MMRs}
\end{figure}

{Using an Earth-mass moon on a circular orbit around a Jupiter-mass planet, we produce Figure \ref{fig:moon_MMRs} to illustrate the locations of the N:1 MMRs for such exomoons.  The host star is assumed to equal $\alpha$ Cen A in terms of its mass (1.133 M$_\odot$) and luminosity (1.519 L$_\odot$).  The white curves in Fig. \ref{fig:moon_MMRs} }are determined through $a_{\rm sat}^N \pm \Delta\sigma_N$ and the color-code denotes the order $N$ of the MMR.  The black regions represent configurations between the MMRs and the light green (solid) curve in Figure \ref{fig:moon_MMRs} traces the first intersection points between adjacent MMRs and can be calculated numerically using a root finding function for $a_{\rm sat}^N - \Delta\sigma_N = a_{\rm sat}^{N+1} + \Delta\sigma_{N+1}$.  An initial condition to the right of the light green curve becomes unstable due to the overlap in the MMRs and this process can be slow due to the eccentricity growth timescale of the satellite.  However, in our four body configuration the planetary eccentricity undergoes oscillations due to secular forcing by the secondary stellar companion.  As a result, the light green curve can shift leftward and more efficiently destabilize prograde satellites. Retrograde satellites are less affected and can survive for a longer timescale at a much larger satellite separation \citep{Henon1970}, but chaotic diffusion within the MMR overlap can lead to instabilities on longer timescales. 

\section{Numerical Determination of Exomoon Stability} \label{sec:numerical}
In this section, we numerically investigate the limits for the stability of satellites in a hierarchical four body problem.  The general setup of our simulation is detailed in Section \ref{sec:setup}.  The retrograde satellite stability limit \citep{Domingos2006} is revisited in Section \ref{sec:revised} to identify the lower critical orbit boundary that was previously overlooked.  Section \ref{sec:orb_stab} examines the limits to exomoon orbital stability using the $\alpha$ Cen AB system.  Finally, the effects of the planetary and stellar tides on Earth-Moon analogs are explored in Section \ref{sec:tides}.

\subsection{Simulation Setup \& Initial Conditions}\label{sec:setup}

N-body simulations provide a direct test of exomoon stability and we use the orbital integration package \texttt{Rebound} \citep{Rein2012,Rein2015} because it is a versatile tool for the evolution of complex N-body systems.  We use the WHFAST and IAS15 integration schemes within \texttt{Rebound} to evaluate the stability of exomoons \citep[i.e.,][]{Rosario-Franco2020} on $10^5$ or $10^6$ year timescales using an initial timestep that is 2.5\% of the exomoon's initial orbital period.  {Since our initial timestep is very small compared to the period of stellar binary, the WHFAST integration scheme maintains sufficient accuracy.  We verified this using the IAS15 adaptive integration scheme at high planetary eccentricity.}  The $10^5$ year timescale is sufficient for the most cases because the typical orbital timescale for an exomoon is much shorter ($\sim$0.1 yr) and the secular timescale for perturbations from the stellar companion is $\sim$10$^4$ years \citep[e.g., $\alpha$ Centauri AB;][]{Quarles2018a,Quarles2018b,Quarles2019}.  Perturbations from the stellar companion on retrograde-orbiting exomoons are weaker, especially in the habitable zone of each star, and require the longer $10^6$ year integrations.  Our definition of stability is for an exomoon that begins on a circular, coplanar orbit to survive for the full integration time.  Unstable conditions are those that cause the exomoon separation $r_{sat}$ to exceed the host planet's Hill radius ($r_{sat} > R_H$) or crash into the host planet ($r_{sat}<R_p$).

Our simulations use the $\alpha$ Centauri AB ($\alpha$ Cen AB) binary to identify the extent of exomoon stability because the stellar masses, separation, and eccentricity are well-known, which is important for our criteria in Section \ref{sec:analytic}.  \cite{Pourbaix2016} provide values for $\alpha$ Cen AB based on analysis of archival data from HARPS, the High Accuracy Radial velocity Planet Searcher at the ESO La Silla 3.6 m telescope.  \citeauthor{Pourbaix2016} find the masses of star A and B are 1.133 M$_\odot$ and 0.972 M$_\odot$, respectively, while the binary orbital semimajor axis and eccentricity are 23.7 au (17.66 arcsec) and 0.524, respectively.  These values update the previous analysis \cite{Pourbaix2002} and find good agreement with estimates from asteroseismology \citep{Lundkvist2014}.  In our simulations, the binary companion is coplanar, begins at periastron, and is nodally aligned (see Table \ref{tab:ICs}).  Our choice for the binary initial phase affects the stability of planetary orbits sufficiently near the stability limit \citep{Quarles2018,Quarles2020}, where the initial separation of the host planets in this work are in a regime that is long-term stable considering a wide range of initial phases for both the binary and planetary orbit \citep{Quarles2020}. 

Since the existence of planets in $\alpha$ Cen AB is currently unknown (much less their atmospheric composition), we choose an initial condition close to the inner edge of the conservative habitable zone \citep{eggl2020habitable}. The habitable zone (HZ) is commonly defined as the region where liquid water could potentially exist on the surface of a rocky planet and depends, among other things, on the properties of the incident radiation and composition of the planet's atmosphere \citep{Kasting1993,Kopparapu2014,eggl2020habitable}. We place the planet at 1.232 au around star A and 0.707 au around star B.  Terrestrial planets on circular orbits at these semimajor axes receive one solar constant's worth of radiative flux annually.
For a planet orbiting $\alpha$ Cen A, the planet's semimajor axis $a_{\rm p}$ is sampled at 1.232 au, 2 au, 2.25 au, or 2.5 au as star A has a greater chance to host planets as far out as $\sim$2--2.5 au \citep{Quarles2016}.  For a planet orbiting $\alpha$ Cen B, the planet's semimajor axis begins at 0.707 au or at 2 au, respectively. In both cases, the planetary orbit starts coplanar ($i_p = 0^\circ$) with the binary orbit, with zero obliquity $\psi_p = 0^\circ$, and is nodally aligned ($\omega_{\rm p}=\Omega_{\rm p} = 0^\circ$).  
Most of the observed circumstellar planets in binary systems are Jupiter-like giants \citep{Schwarz2016}.  For the planetary mass M$_p$ and radius R$_p$, we use values identical to Jupiter's in our orbital stability simulations.  Prior studies of the satellites in the solar system \citep{Estrada2006,Canup2006} developed formation pathways akin to the terrestrial planets, but for a circumplanetary disk.  \cite{Canup2006} suggested that in-situ formation of moons from circumplanetary disks around giant planets are limited to $\sim$10$^{-4}$ M$_p$ due to the estimated flux of infall material from the circumstellar disk, but this prediction awaits confirmation from exoplanet observations.  We use an Earth-mass exomoon M$_{\rm sat}$ in our simulations due to the potential of tidal capture through scattering events and the flux of smaller infall material could be larger than was for the solar system due to the formation environment of the binary stars.  However, we switch to an Earth-Moon analog for the planet-moon system when considering the possible effects of tidal evolution (see Section \ref{sec:tides}).  This is motivated by observational constraints for $\alpha$ Cen AB \citep{Zhao2018} and the comparisons that we make with the solar system. 

For the exomoons, we explore prograde and retrograde orbits because both are ubiquitous within the solar system \citep{Jewitt2007}, where there could be a higher incidence of scattering during the early stages of planet formation \citep{Quintana2002,Quintana2007,Haghighipour2007}.  Since we are investigating long-lived exomoons, we expect that the tidal dissipation in the host planet to efficiently align the exomoon's orbit with the planetary equator (i.e., coplanar with the planetary and binary orbits) and circularize the exomoon's orbit.  As a result, our numerical model begins the exomoons on circular ($e_{sat} = 0$) orbits that are nodally aligned $\omega_{\rm sat}=\Omega_{\rm sat} = 0^\circ$ with the planet and binary orbits.  We vary the planet-satellite separation $a_{sat}$ as a fraction of the planet's Hill radius starting at 0.25 $R_{\rm H}$ and up to 0.7 $R_H$ in 0.01 $R_{\rm H}$ steps, where the 0.25 $R_{\rm H}$ is just interior to planet-satellite MMRs and 0.7 $R_{\rm H}$ is the theoretical limit for stability for retrograde small bodies in the circular restricted three body problem \citep{Quarles2020}.  

Following \cite{Rosario-Franco2020}, we evaluate 20 initial phases $f_{\rm sat}$ that are randomly drawn between 0$^\circ$--180$^\circ$ from a uniform distribution for each exomoon semimajor axis $a_{\rm sat}$.  When constructing our initial conditions, we use the above prescribed orbital elements and convert pairwise (e.g., [A,(p,sat),B]) to the Cartesian positions and velocities.  The center of mass between the planet and satellite typically resides within the planet and thus our results are indistinguishable from a Jacobian construction of initial conditions.  Table \ref{tab:param_rng} summarizes the parameter ranges that we explore for the orbital stability simulations.

\begin{table*}
	\centering
	\caption{Parameters used for the masses, luminosities, and orbit of $\alpha$ Cen AB {{in a reference plane aligned with the eccentricity and angular momentum vectors of the binary}}.}
	\label{tab:ICs}
	\begin{tabular}{cccccccccc} 
		\hline
		 M$_{\rm A}$ (M$_\odot$) & L$_{\rm A}$ (L$_\odot$) & M$_{\rm B}$ (M$_\odot$) & L$_{\rm B}$ (L$_\odot$) & $a_{\rm bin}$ (au) & $e_{\rm bin}$ & $i_{\rm bin}$($^\circ$) & $\omega_{\rm bin}$($^\circ$) & $\Omega_{\rm bin}$($^\circ$) & $f_{\rm bin}$($^\circ$)\\
		\hline
		1.133 & 1.519 & 0.972 & 0.500 & 23.78 & 0.524 & 0 & 0 & 0 & 0\\
		\hline
	\end{tabular}
\end{table*}

\begin{table}
	\centering
	\caption{Parameter ranges used in orbital stability simulations.}
	\label{tab:param_rng}
	\begin{tabular}{ccc} 
		\hline
		 parameter & unit &  range/value\\
		\hline
		M$_{\rm p}$ & M$_\oplus$ & 318 \\
		$a_{\rm p}$& au & $\sqrt{L_\star}$, 2, 2.25, 2.5\\
		$e_{\rm p}$&  & 0.00 -- 0.60 \\
		M$_{\rm sat}$ & M$_\oplus$ & 1 \\
		$a_{\rm sat}$ & R$_{\rm H}$ & 0.25 -- 0.70\\
		$i_{\rm sat}$ & $^\circ$ & 0, 180 \\
		$f_{\rm sat}$ & $^\circ$ & 0 -- 180 \\
		\hline
	\end{tabular}
\end{table}

\subsection{Revised Retrograde Stability Limit}\label{sec:revised}
Previous studies \citep{Henon1970,Innanen1979,Hamilton1991,Domingos2006} show that the extent of stable orbits for retrograde-orbiting moons is substantially larger than for a prograde moon.  \cite{Domingos2006} explored how the stability limit for retrograde satellites varied with the planet and satellite eccentricity.  However, the simulations by \citeauthor{Domingos2006} evaluated only a single initial phase ($f_{\rm sat} = 0^\circ$) to formulate the stability boundary.  As a result, the empirically determined retrograde stability limit was actually the upper critical orbit \citep{Dvorak1986}.  This distinction to their study was discussed recently for the case of prograde-orbiting satellites \citep{Rosario-Franco2020}.  Moreover, \cite{Quarles2020} numerically expanded the Jacobi constant stability criterion \citep{Eberle2008} to retrograde orbits and determined that the retrograde stability limit $a_c \approx 0.7$ R$_H$ for the $\mu = 10^{-3}$ case studied by \cite{Domingos2006}.  

\begin{figure}
	\includegraphics[width=\columnwidth]{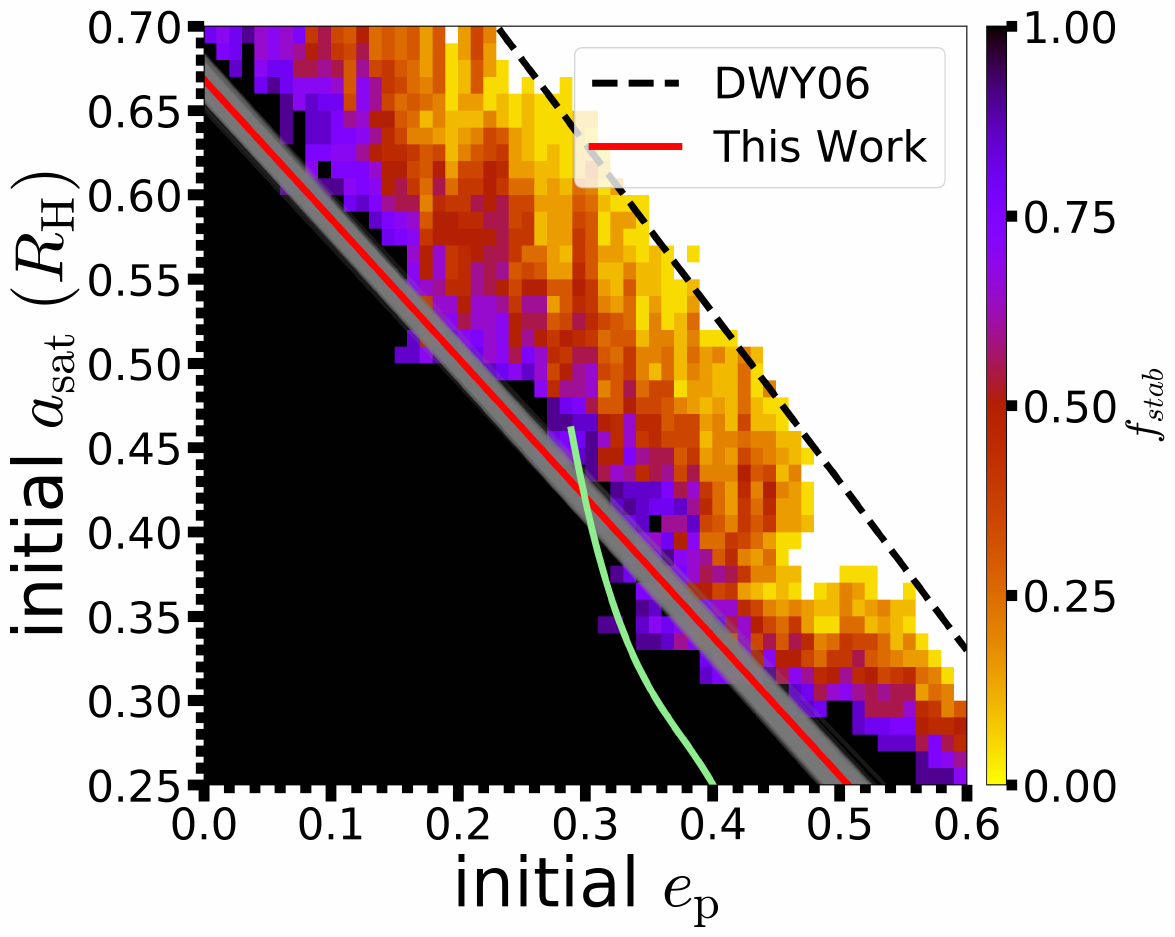}
    \caption{Numerical estimates of the stability for retrograde orbiting exomoons in \emph{single star systems} in terms of the stability fraction $f_{\rm stab}$ (color-coded), \textit{initial} host planet eccentricity $e_{\rm p}$, and exomoon separation $a_{\rm sat}$ in units of the planetary Hill radius $R_{\rm H}$.  The white cells denote unstable initial conditions, where all 20 trials terminated before $10^5$ yrs.  The black (dashed) curve marks the upper critical orbit \citep[][; DWY06]{Domingos2006} and the red (solid) curve is the lower critical orbit determined by this work, where the gray (solid) curves show the uncertainty.  The light green (solid) curve illustrates the boundary from Fig. \ref{fig:moon_MMRs}, where MMR overlap can occur.  }
    \label{fig:moon_retro}
\end{figure}

We perform simulations of retrograde {Earth-mass} exomoons hosted by Jupiter-mass planet that orbits 1 au from its Solar-mass host star (see Section \ref{sec:setup}).  The exomoon begins on a circular orbit, where the host planet's eccentricity is varied from 0--0.6 in 0.01 steps.  We test initial exomoon semimajor axis values starting within 0.25--0.70 R$_{\rm H}$, where the outer limit is the determined from the Jacobi constant criterion \citep{Eberle2008,Quarles2020}.  The planet and satellite orbits begin nodally aligned, where we evaluate 20 random values for the initial satellite orbital phase $f_{\rm sat}$.  Figure \ref{fig:moon_retro} illustrates the relation between the initial satellite separation $a_{\rm sat}$ and the planetary eccentricity $e_p$ using a color-code for the fraction of initial phases $f_{\rm stab}$ that are stable for the full integration timescale of $10^5$ years.  {Since these simulations lack a stellar companion the longer 10$^6$ timescale described in Section \ref{sec:setup} is unnecessary, but will be used for retrograde systems in Section \ref{sec:orb_stab}.}

The white cells in Figure \ref{fig:moon_retro} mark when all 20 trials either collide with or are stripped from the host planet.  The black cells mark when all 20 trials are stable, where the colored cells show the transition.  The black (dashed) curve marks the prediction from the fitting formula by \cite{Domingos2006}, where the red (solid) curve denotes our fitted stability boundary using a linear function in the planetary eccentricity in units of the Hill radius $R_{\rm H}$, 
\begin{equation}
    a_{\rm sat}^{\rm crit} = C_1(1-C_2e_p)\label{eq:retrofit}
\end{equation}
and the gray curves show the uncertainty in our estimate. Equation \ref{eq:retrofit} has a similar structure as Equation \ref{eq:stab} where the planetary eccentricity is used directly (instead of the maximum eccentricity) because the planetary eccentricity is nearly constant in time for these Sun-Jupiter-satellite systems with negligible tidal interactions.  Figure \ref{fig:moon_retro} demonstrates that the boundary from \cite{Domingos2006} and our boundary represents the upper and lower critical orbits \citep{Dvorak1986}, respectively.  Our stability boundary recovers the limit set by the Jacobi constant criterion \citep{Eberle2008,Quarles2020} for circular planetary orbits.  The red (solid) curve is fit by the coefficients: $C_1 = 0.668 \pm 0.006 R_H$ and $C_2 = 1.236 \pm 0.019$.  Using our outer stability limit coefficient $C_1$ is important because it significantly affects the limiting satellite mass (M$_{\rm sat} \propto C_1^{13/2}$) through tidal migration \citep{Barnes2002}.

\subsection{Orbital stability in $\alpha$ Centauri AB} \label{sec:orb_stab}
Planets orbiting either component in a stellar binary experience strong perturbations, but the perturbation strength scales with the pericenter distance of the secondary star and the initial semimajor axis of the planet \citep[e.g.,][]{David2003,Quarles2018}.  We perform numerical simulations (see \S\ref{sec:setup}) of prograde and retrograde orbiting satellites using the $\alpha$ Cen AB binary and consider host planets whose orbits begin at the inner edge of their respective HZs (1.232 au or 0.707 au) or closer to their respective stability limits ($a_{\rm p}\sim$2--2.5 au).  $\alpha$ Cen AB is representative of moderately wide binary systems \citep{Raghavan2010,Moe2017} in terms of its eccentricity and binary orbital period, which makes it important as a case study.  The distribution of binary eccentricity and orbital period is quite broad and we explore the possible impacts on our results in Section \ref{sec:monte_carlo}.

\begin{figure}

	\includegraphics[width=\columnwidth]{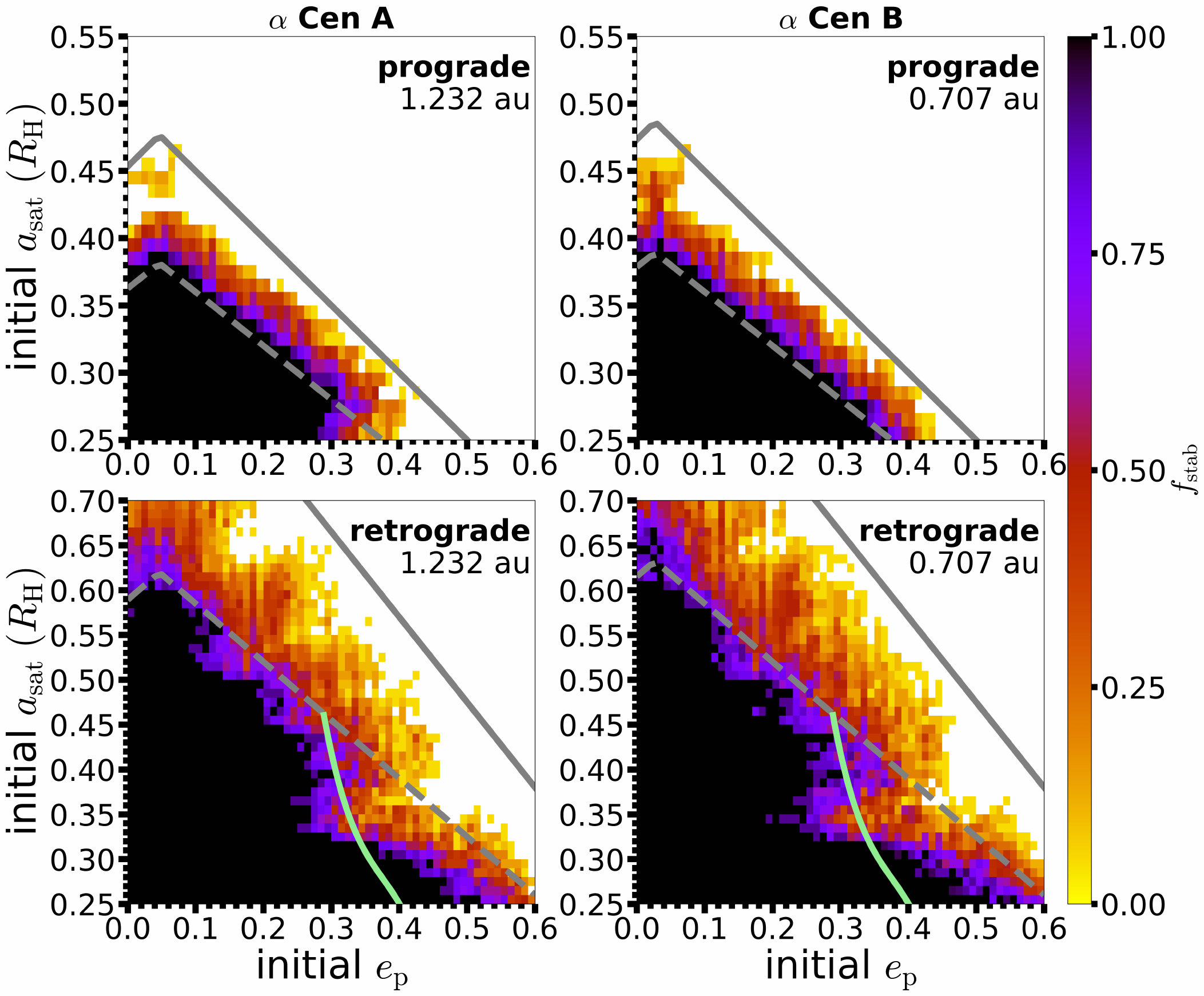}
    \caption{Similar to Figure \ref{fig:moon_retro}, but the Jupiter-mass planet orbits either star in $\alpha$ Cen AB at the inner edge of its respective HZ with an Earth-mass moon. The gray (solid and dashed) curves mark the upper and lower critical orbit boundaries (see Equation \ref{eq:bin_stab}).  The light green (solid) curve illustrates the boundary from Fig. \ref{fig:moon_MMRs}, where MMR overlap can occur.}
    \label{fig:moon_hz}
\end{figure}

Starting with planets and moons in the respective HZ of their host star in $\alpha$ Cen AB, we numerically determine the outer stability boundary for satellites (Figure \ref{fig:moon_hz}).  We also plot two gray curves that represent the upper or lower critical orbit using the formalism derived in Section \ref{sec:analytic} in units of the planetary Hill radius relative to the host star (Table \ref{tab:param_rng}),
\begin{equation} \label{eq:bin_stab}
    a_{\rm sat}^{\rm crit} = C_{\rm sat}(1-\epsilon_{\rm p} - \eta_{\rm p}) R_{\rm H}
\end{equation}
where $C_{\rm sat}$ is a scale factor for $R_H$ in Equation \ref{eq:stab}. The coefficient $C_{\rm sat}$ corresponds to values determined for exomoons in single star systems \citep{Domingos2006,Rosario-Franco2020} for prograde and retrograde orbits.  For the solid gray curve representing the upper critical orbit, $C_{ \rm sat}$ is 0.5 (prograde) or 0.95 (retrograde).  For the dashed gray curve representing the lower critical orbit, $C_{\rm sat}$ is 0.4 (prograde) or 0.65 (retrograde).  {Due to the absolute value imposed in {Equation \ref{eq:efree}}, the gray curves have a peak at the forced eccentricity, $\epsilon_{\rm p}$.}  Similar to Figure \ref{fig:moon_retro}, the color-code in Figure \ref{fig:moon_hz} illustrates the transition between stable and unstable orbits.  In each panel, the solid curve encapsulates nearly all the potentially stable cells, while the dashed curve marks the more conservative inner boundary.  The curves fit reasonably well because the eccentricity growth of the satellite, which is the main driver for instabilities, grows secularly in this regime.  Moderately eccentric planets ($e_{\rm p} \geq 0.3$) undergo eccentricity oscillations and MMR overlap causes significant eccentricity growth for the satellite.  The light green (solid) curves in Figure \ref{fig:moon_hz} (lower panels) mark the boundary where MMR overlap begins in the host star-planet-satellite system. Those curves shift left and right as the system evolves. 

\begin{table}
	\centering
	\caption{Stability Coefficients $ C_{\rm sat}$ to determine the Upper and Lower Critical Orbits for Exomoons in $\alpha$ Cen AB.}
	\label{tab:stab_coeff}
	\begin{tabular}{ccc} 
		\hline
		  & prograde & retrograde\\
		\hline
		LCO & 0.4 & 0.65\\
		UCO & 0.5 & 0.95\\
		\hline
	\end{tabular}
\end{table}

There are a handful of stellar binaries with a separation of $\sim$20 au that are known to host Jovian exoplanets near the stability limit (e.g., HD 196885AB, \cite{Chauvin2007,Correia2008,Fischer2009,Chauvin2011}; $\gamma$ Cephei AB \cite{Hatzes2003,Torres2007}) with extensive dynamical studies focusing on the stability of the planet \cite{Thebault2011,Giuppone2011,Giuppone2012,Satyal2013,Satyal2014}.  We perform numerical simulations of similar conditions for a putative Jupiter-like exoplanet orbiting either star in $\alpha$ Cen AB at 2, 2.25, or 2.5 au from the host star to investigate the stability of moons under more significant perturbations from the stellar companion.  \cite{Zhao2018} may have already excluded Jupiter-sized planets at these distances using archival radial velocity measurements, but our results are scaled by the planetary Hill radius and would still be applicable for lower mass planet-satellite pairs as long as M$_{\rm sat}\ll  $ M$_{\rm p}$.  

\begin{figure}
	\includegraphics[width=\columnwidth]{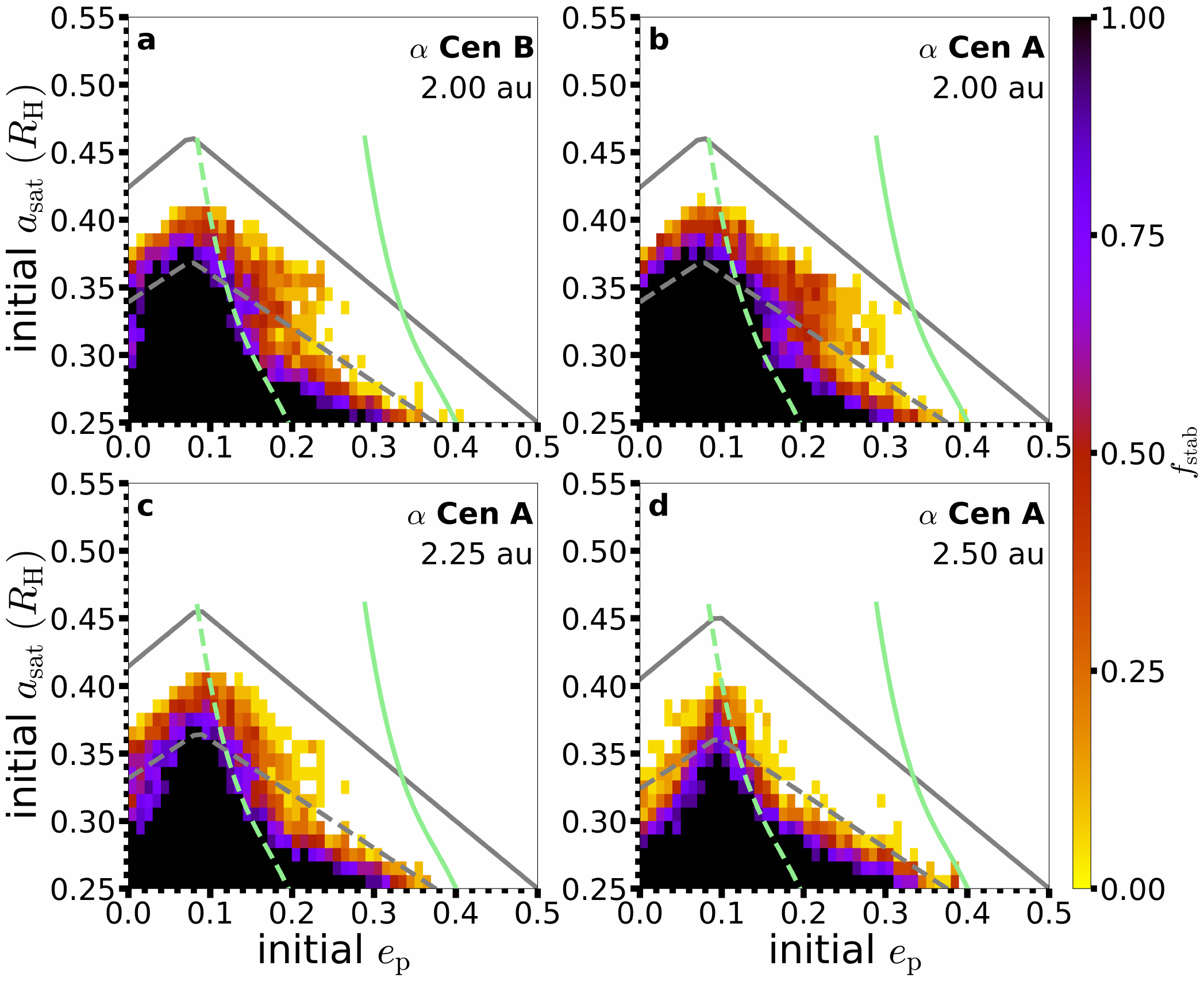}
    \caption{Similar to Figure \ref{fig:moon_hz}, but the Jupiter-mass planet orbits either star in $\alpha$ Cen AB closer to its stability limit and its Earth-mass moon begins on a \textit{prograde} orbit. Specifically, the initial host planet semimajor axis is indicated in each of the respective panels. The gray (solid and dashed) curves mark the upper and lower critical orbit boundaries (see Equation \ref{eq:bin_stab}).  The light green (solid) curve illustrates the boundary from Fig. \ref{fig:moon_MMRs}, where MMR overlap can occur and the light green (dashed) curve denotes the constraint on stability due to {the evolution of the planet and moon eccentricity that alters the location of the MMR overlap region (see Eqn. \ref{eq:MMR})}.}
    \label{fig:moon_bin_stab}
\end{figure}

Figure \ref{fig:moon_bin_stab} illustrates our results for prograde orbiting satellites and is similar to Figure \ref{fig:moon_hz} (top row).  Satellites can be stable at larger separations when the host planet orbits its star at the forced eccentricity $\epsilon_{\rm p}$ (see \S\ref{sec:analytic}) and is apsidally aligned ($\omega_p = \omega_{bin}$) with the binary orbit \citep{andrade2017secular,Quarles2018}.  Apsidal alignment is important because it limits the degree of precession for the planetary orbit and minimizes the planet's eccentricity oscillations, where the maximum planetary eccentricity $e_{\rm p}$ can reach $\sim$0.2 when starting from a circular orbit \citep{Quarles2018}. { For an eccentric planetary orbit in the HZ of $\alpha$ Cen AB, $e_{\rm p}$ can also vary by $\sim$0.2 over a secular cycle \citep{Quarles2018}, but the magnitude depends on the proximity to the forced eccentricity.}  At the maximum of the planetary eccentricity oscillation, the truncation of the planetary Hill radius is more significant and significant overlap with MMRs can occur, where an ejection of the satellite becomes more probable.  

\begin{figure}
	\includegraphics[width=\columnwidth]{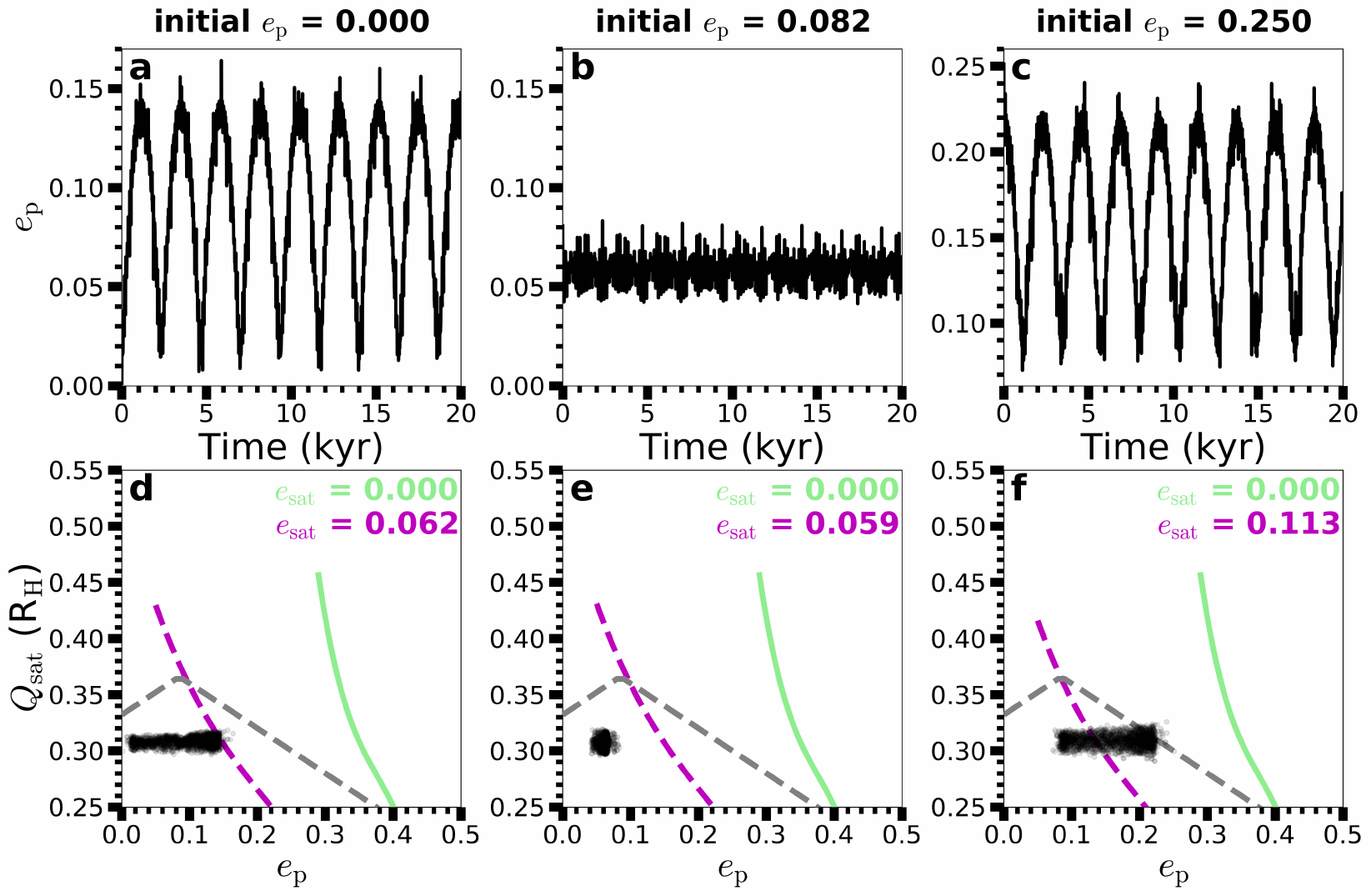}
    \caption{{Short-term evolution of the eccentricity of the planetary orbit (top) and the apocenter distance of the moon (bottom) using initial conditions sampled from Fig. \ref{fig:moon_bin_stab}c, where the initial semimajor axis for each moon is  0.3 R$_{\rm H}$ and initially circular.  Panels a--c demonstrate the secular evolution of the host planet's eccentricity $e_{\rm p}$ over 20 kyr.  Panels d--f illustrate the evolution of the moon's apocenter distance $Q_{\rm sat}$ in response to the planetary eccentricity evolution (panels a--c, respectively).  The gray (dashed) curve marks lower critical orbit boundary, where the light green (solid) and magenta (dashed) curves denote the boundary for MMR overlap for the minimum and maximum satellite eccentricity, respectively.  In panels d \& f the moon eventually becomes unstable as a result of entering the MMR overlap region, whereas the moon can remain stable in panel e due to a lower amplitude of variation in the host planet's eccentricity (panel b).}}
    \label{fig:moon_sec_evol}
\end{figure}

\begin{figure}
	\includegraphics[width=\columnwidth]{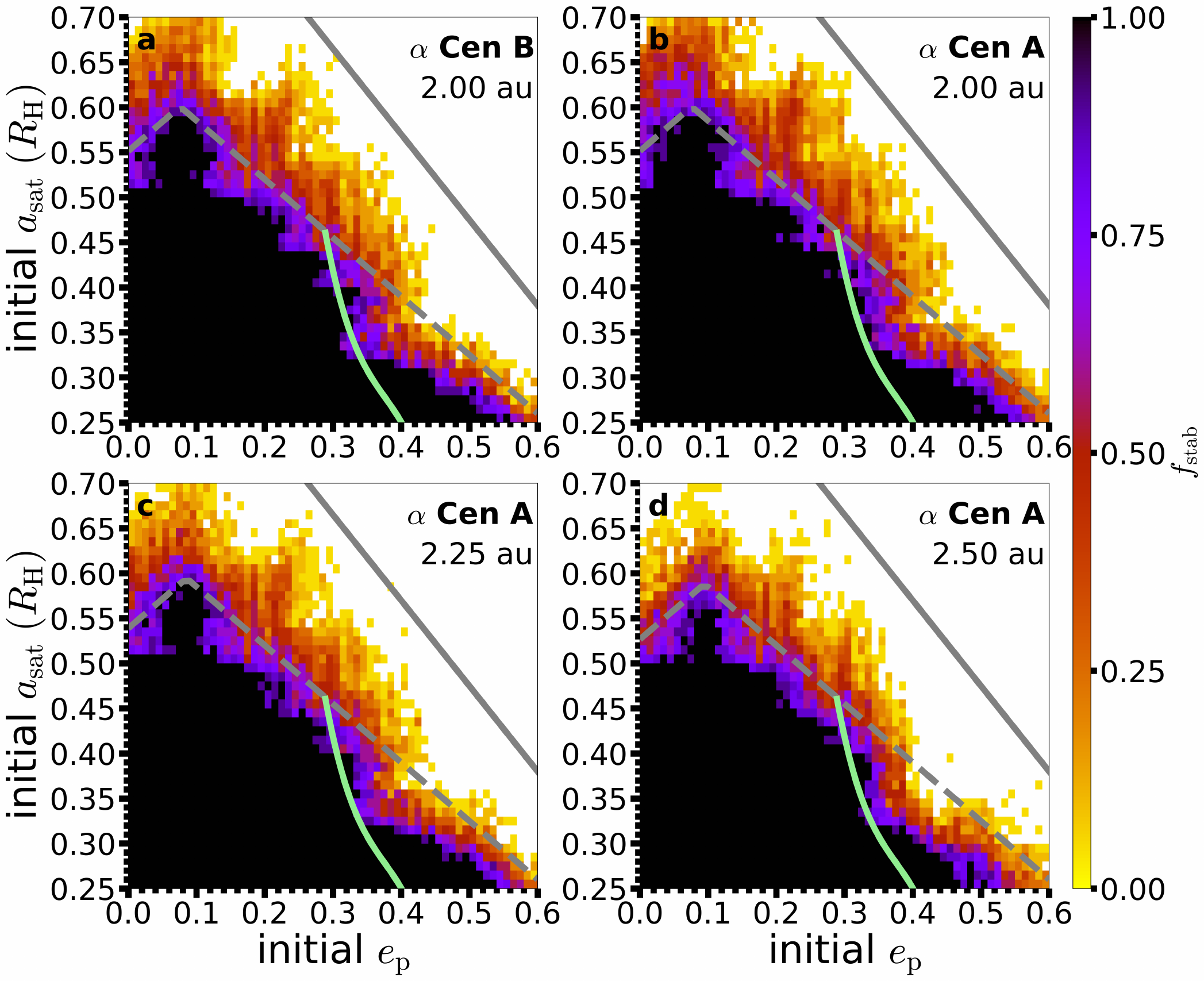}
    \caption{Similar to Figure \ref{fig:moon_bin_stab}, but the Earth-mass moon begins in \textit{retrograde} relative to the orbit of the host planet. The gray (solid and dashed) curves mark the upper and lower critical orbit boundaries (see Equation \ref{eq:bin_stab}).  The light green (solid) curve illustrates the boundary from Fig. \ref{fig:moon_MMRs}, where MMR overlap can occur.}
    \label{fig:moon_bin_retro}
\end{figure}

Figure \ref{fig:moon_bin_stab} demonstrates this effect through: 1) the gray curves marking the upper and lower critical satellite orbits and 2) the light green curves denoting the boundary for overlapping MMRs.  The dashed (light green) curves represent a shift of the MMR based curves (solid) to smaller initial planetary eccentricities by about $\sim$0.2, which is representative of the maximum oscillation of the planetary eccentricity that permits planetary stability from previous studies \citep{Quarles2016,Quarles2018}. {Figure \ref{fig:moon_sec_evol} illustrates the planetary eccentricity oscillations (Fig. \ref{fig:moon_sec_evol}a--c) for three different initial values of $e_{\rm p}$, where the remaining initial conditions are drawn from Fig. \ref{fig:moon_bin_stab}c and $a_{\rm sat}$ begins at 0.3 R$_{\rm H}$.  Figs. \ref{fig:moon_sec_evol}a \& \ref{fig:moon_sec_evol}c show a higher variation due to their relative distance from the planetary forced eccentricity ($\epsilon_{\rm p} \sim 0.08$ when $a_{\rm p} =$ 2.25 au).  Figure \ref{fig:moon_sec_evol}d--f demonstrate the evolution of the satellite's apocenter $Q_{\rm sat}$ (black dots) in response to the planet's eccentricity variation, where the gray (dashed) curve marks the lower critical orbit boundary, light green (solid) curve denotes the boundary for MMR overlap when $e_{\rm sat}= 0$, and the magenta (dashed) shows the shift of the MMR overlap boundary at the maximum satellite eccentricity.  The boundary for MMR overlap depends on $e_{\rm p}$ and $e_{\rm sat}$ (see Eqn. \ref{eq:MMR}), where increases in $e_{\rm sat}$ lead to shifts in the MMR overlap boundary to lower $e_{\rm p}$ values  and increases in $e_{\rm p}$ (relative to $\epsilon_{\rm p}$) correlates with a higher likelihood of reaching the new boundary.  Instability can then ensue once the satellite enters the MMR overlap region, but this process is chaotic \citep{Mudryk2006}.}  Figure \ref{fig:moon_bin_retro} illustrates similar calculations as Fig. \ref{fig:moon_bin_stab}, but for retrograde orbiting satellites.  The overall area is greater due to the enhancement to stability from retrograde orbits and the possible truncation due to planetary eccentricity oscillations is less severe.  Even though more of the parameter space is stable for retrograde satellites, the maximum satellite eccentricity is also larger and the possible existence of multiple satellites would constrain the parameter space further \citep{Giuppone2013}.

\subsection{Tidal migration lifetimes in $\alpha$ Centauri AB} \label{sec:tides}

In our solar system, the lifetime of moons is significantly constrained by tidal interactions \citep{Barnes2002, Sucerquia2019, lainey2020resonance}. While constraints have been placed on the long-term tidal evolution of exomoons in single stellar systems \citep{Sasaki2012, Sasaki2014}, this has yet to be established for exomoons in stellar binary systems.  To obtain a full picture of orbital stability of exomoons in stellar binary systems, it is necessary to consider the contribution of planetary and stellar tides.  In this section, we apply a secular constant time lag (CTL) tidal model and evaluate the migration lifetimes of exomoons within the HZs of each stellar binary component in $\alpha$ Cen AB. 

Analyses of observational surveys \citep{Zhao2018} have largely excluded Jupiter-mass exoplanets from the HZs of $\alpha$ Cen AB, where the detection thresholds are 53 M$_\oplus$ and 8.4 M$_\oplus$ for star A and B, respectively.  Additionally, {the low tidal time lag in Jupiter-like planets ($k_{\rm 2,J}\Delta t_{\rm J} \sim 10^{-2}$ s or $Q_{\rm J} \sim 10^6$; see \cite{Leconte2010}) greatly reduces the outward migration of satellites and prevents us from placing a strong constraint using tidal migration on possible exomoons}, although recent results for Saturn suggest substantially faster migration rates could be possible \citep{lainey2020resonance}. {More complicated models for the tidal dissipation in gas dominated exoplanets could be employed \citep[e.g.,][]{Guenel2014,AlvaradoMontes2017}, but such considerations are beyond the scope of this work.}
As a consequence we use an Earth-Moon analog in our tidal model instead, due to its potential to place meaningful constraints on exomoons in $\alpha$ Cen AB, and because the Earth-Moon system has tidal parameters that are better understood.  Similar to section \ref {sec:orb_stab}, the planetary orbits are sampled from the inner edge of their HZ, but values for exomoon's semimajor axis begin at 8.64 $R_\oplus$ (three times the Roche radius of an Earth-Moon analog). 

\begin{figure}
	\includegraphics[width=\columnwidth]{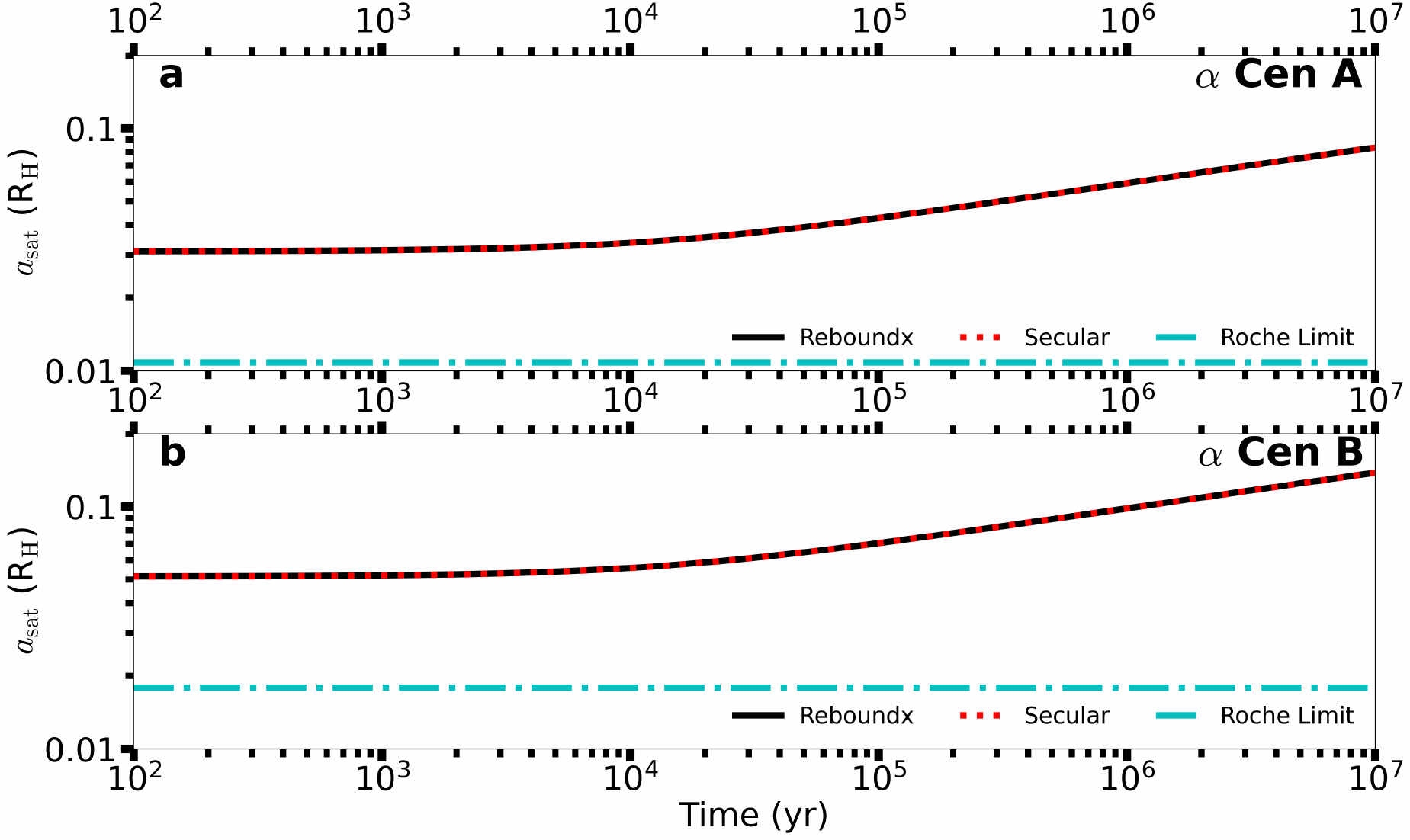}
    \cprotect\caption{Tidal evolution for {10 Myr} of an Earth-Moon analog near the inner edge of the conservative HZ of each star in $\alpha$ Cen AB using a constant time lag tidal model \citep{HUT1981}.  The N-body approach (solid black) calculates the tidal evolution due to the instantaneous torques { (using the \verb|tides_constant_time_lag| module \citep{Baronett2021} in \texttt{reboundx})} and the secular (dotted red) method approximates the evolution averaged over an orbit.  The satellite begins at 0.03--0.05 R$_{\rm H}$ or 3$\times$ the Roche limit (dash-dot cyan).}
    \label{fig:Nbody_comp}
\end{figure}

We implement a constant time lag secular model \citep{Leconte2010, HUT1981} and evaluate the tidal evolution up to 10 Gyrs. This model calculates the secular changes to the semimajor axes ($a_{\rm p}$ and $a_{\rm sat}$), eccentricities ($e_{\rm p}$ and $e_{\rm sat}$), and mean motion ($n_{\rm p}$ and $n_{\rm sat}$) averaged over an orbit.  The model is scaled by the tidal Love number $k_2$ and the time lag $\Delta t$, where the latter is proportional to $(nQ)^{-1}$ in the constant phase lag (i.e., constant $Q$) tidal models \citep{Leconte2010,Piro2018}.  A reasonable time lag for the Earth-Moon system is $\sim$100 s, and we assume $\Delta t_{\rm p} = 100$ s, unless stated otherwise.  The initial rotation period of the host planet is 5 hours, which is consistent with expectations from terrestrial planet formation \citep{Kokubo2007}.  The external perturbations from the secondary star on the exomoon are weak compared to the planetary gravitational interactions and a tidal model considering the interactions between the host star-planet-moon are sufficient.  We use { a new module called \verb|tides_constant_time_lag| from \texttt{reboundx} \citep{Baronett2021}} to test this assumption, which is based upon the well-established code \texttt{mercury-T} \citep{Bolmont2015}.  { Our implementation of the \verb|tides_constant_time_lag| module} evaluates the instantaneous torques at each timestep { using the \texttt{ias15} integrator} and the N-body simulations are evolved for 10 Myrs, which is an appropriate timescale to see the effects on the satellite orbit due to tides.  {At the time of this writing, the} \verb|tides_constant_time_lag| {module does not evolve the changes to primary body's spin due to outward tidal migration of a secondary body.  Therefore, we update the spin rate of the primary body periodically using interpolated values of the host planet's spin rate from the afore mentioned secular model.}  Figure \ref{fig:Nbody_comp} illustrates the results from \texttt{reboundx} (solid black) and the corresponding secular model (dashed red), where the two approaches are in agreement to a high degree.  In both models, the exomoon's eccentricity grows at similar rates and orbital energy exchanges between the planet and moon become possible.  Since the N-body method is more computationally expensive (12 CPU-days), we use the secular model for the remainder of this section.

\begin{figure}
	\includegraphics[width=\columnwidth]{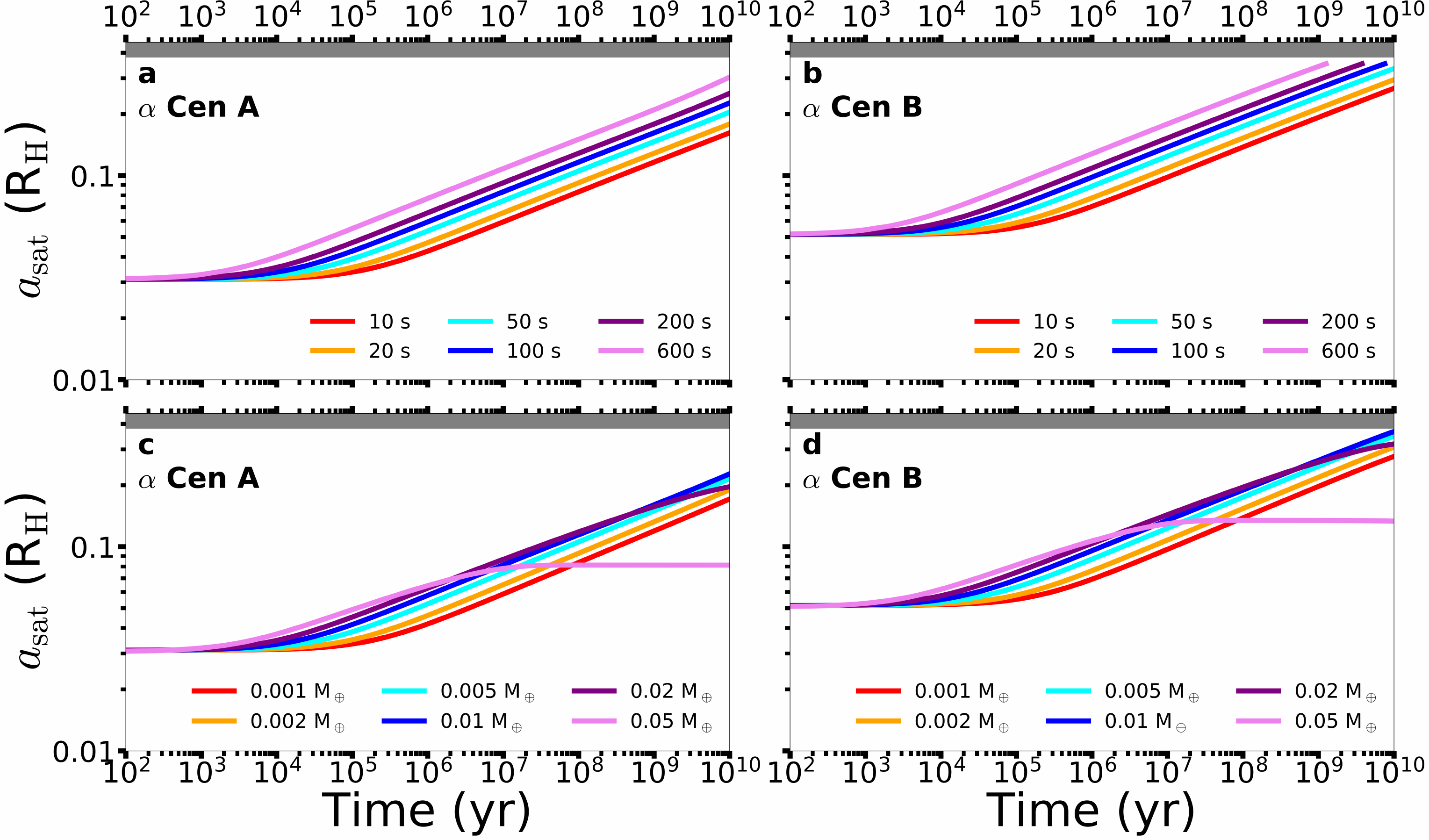}
    \caption{Evolution of tidal models for an Earth-analog planet and putative satellites in $\alpha$ Cen AB varying the time lag (a \& b; M$_{\rm sat} = { 0.0123}$ M$_\oplus$) or the satellite mass (c \& d; $\Delta t_{\rm p} = 100$ s).  The host planet begins near the inner edge of the conservative HZ at its forced eccentricity (see Equation \ref{eq:eforced}), where the satellite starts on a low eccentricity ($e_{\rm sat} = 10^{-3}$) orbit at 3$\times$ the Roche limit.  The gray region marks the unstable region for prograde orbits (see Figure \ref{fig:moon_hz}). }
    \label{fig:param_comp}
\end{figure}

The outward migration of a satellite varies with both the assumed time lag and with the mass of the satellite.  We evaluate the outward tidal migration using a broad range in the time lag (10--600 s) and the satellite mass (0.001--0.05 M$_{\rm p}$) over a $10^{10}$ year timescale (i.e., main sequence lifetime of a G dwarf).  Figure \ref{fig:param_comp} shows variation in outcomes due to these parameters for an Earth-Moon analog orbiting either stellar host in $\alpha$ Cen AB.  The color-code delineates the assumed parameter in each panel.  Figures \ref{fig:param_comp}a and \ref{fig:param_comp}b evolve a Moon-like satellite (0.0123 M$_\oplus$).  Satellites are stable for the full range of time lags considered for $\alpha$ Cen A.  But, exomoons can escape after 1 Gyr when their host planet orbits $\alpha$ Cen B for a time lag $\gtrsim$100 s.  For $\alpha$ Cen B, the inner edge of the HZ is closer to the host star and the tidal de-spinning of the host planet is faster, which can accelerate the outward migration.  Figures \ref{fig:param_comp}c and \ref{fig:param_comp}d consider a range of satellite masses, where the time lag is held fixed at 100 s.  There is a smaller range of outcomes when the satellite mass is varied, where larger planet-satellite mass fractions ($\gtrsim$0.02 M$_{\rm sat}/$M$_{\rm p}$) shorten the de-spinning timescale for the host planet.  Once the host planet is de-spun, the migration changes direction and the satellite begins to fall towards the planet.  In this case, the stellar tide on the host planet is weak enough so that the infall rate is slow as is shown by the flattening of the 0.05 M$_\oplus$ curves in Figures \ref{fig:param_comp}c and \ref{fig:param_comp}d.  If the host planet's orbit was $\sim$0.5$\times$ smaller, then a putative satellite would collide with the host planet within 10 Gyrs \citep[e.g.,][]{Sasaki2012,Sasaki2014}.

Previous studies have historically used tidal migration to place upper limits on satellite masses using a constant phase lag (i.e., constant $Q_{\rm p}$) tidal model \citep{Goldreich1966,Barnes2002,Domingos2006}.  However, these efforts are usually limited to cases where the satellite mass is much smaller than the planet (i.e., M$_{\rm sat}/$M$_{rm p} \ll 1$) so that the satellite mass is neglected in the formulation of the satellite's mean motion $n_{\rm sat}$.  Additionally, the final expression is determined in terms of a fraction of the planet’s Hill radius $f$ \citep[see their Equation 8][]{Barnes2002}.  In the constant phase lag model, the satellite's semimajor axis changes via the following differential equation \citep{Murray1999}:

\begin{equation}\label{eq:adot_Q}
    \dot{a}_{\rm sat} = \frac{3 k_{\rm 2,p}}{Q_{\rm p}}
    \sqrt{\frac{GM_{\rm p}}{a_{\rm sat}}}
    \frac{M_{\rm sat}}{M_{\rm p}}
    \sqrt{1+\frac{M_{\rm sat}}{M_{\rm p}}}
    \left(\frac{R_{\rm p}}{a_{\rm sat}}\right)^5,
\end{equation}
where $G$ is the Gravitational constant, R$_{\rm p}$ is the planetary radius, $k_{\rm 2,p}$ is the planetary Love number, and $Q_{\rm p}$ is the tidal quality factor.  When the initial planetary rotation period is sufficiently short \citep{Piro2018}, the satellite migrates past the stability limit (Equation \ref{eq:bin_stab}) on a timescale $T$ that is determined through the integral of Equation \ref{eq:adot_Q} resulting in:
\begin{equation}
T = \frac{2}{13}\left(a_{\rm f}^{13/2} - a_{\rm o}^{13/2} \right)\frac{Q_{\rm p}}{3k_{\rm 2,p}R_{\rm p}^5} \left(\sqrt{GM_{\rm p}\left(1+\frac{M_{\rm sat}}{M_{\rm p}}\right)} \frac{M_{\rm sat}}{M_{\rm p}} \right )^{-1}
\label{eq:tidalT}
\end{equation}
where $a_{\rm f}$ and $a_{\rm o}$ are the final and initial semimajor axes of the satellite, respectively.  For moons that form close to the host planet, the term proportional to $a_{\rm o}$ can be neglected because $a_{\rm f} \gg a_{\rm o}$.  To determine the limiting mass ratio, we make the substitution $x_{\rm m} = M_{\rm sat}/M_{\rm p}$ and set $a_{\rm f}$ equal to $f$R$_{\rm H}$, where $f=C_{\rm sat}(1-\epsilon_{\rm p} - \eta_{\rm p})$ from Equation \ref{eq:bin_stab}. Rewriting $R_H$ in terms of $x_m$ as  $R_{\rm H}=a_{\rm p}[(1+x_{\rm m})M_{\rm p}/(3M_\star)]^{1/3}$ and inserting $a_{\rm f}$ into Equation \ref{eq:tidalT} yields the following expression after straight forward algebraic manipulation:
\begin{equation} \label{eq:masslim}
    \frac{x_{\rm m}}{(1+x_{\rm m})^{5/3}} \leq \frac{2}{13}\left[ fa_{\rm p} \left(\frac{M_{\rm p}}{3M_\star}\right)^{1/3}\right ]^{13/2} \frac{Q_{\rm p}}{3k_{\rm 2,p}TR_{\rm p}^5\sqrt{GM_{\rm p}}}.
\end{equation}
Equation \ref{eq:masslim} represents the upper limit in the mass ratio $x_{\rm m}$ between the satellite and the planet for a given configuration after a set time $T$.
For the range of parameters considered in this work Equation \ref{eq:masslim} has only one real root which is most easily found numerically using a root-finding algorithm, such as the \verb|root_scalar| function in \texttt{numpy} \citep{Harris2020}. Equation \ref{eq:masslim}  reduces to Equation 8 from \cite{Barnes2002} in the low mass ratio limit (i.e., $x_{\rm m} \ll 1$).  As pointed out by \cite{Piro2018}, Equation \ref{eq:masslim} will hold as long as sufficient spin angular momentum in the host planet is available for the outward migration.  Once the planetary rotation synchronizes with the satellite's mean motion, the outward migration stops and inward migration towards the Roche limit can occur.  

\begin{figure}
	\includegraphics[width=\columnwidth]{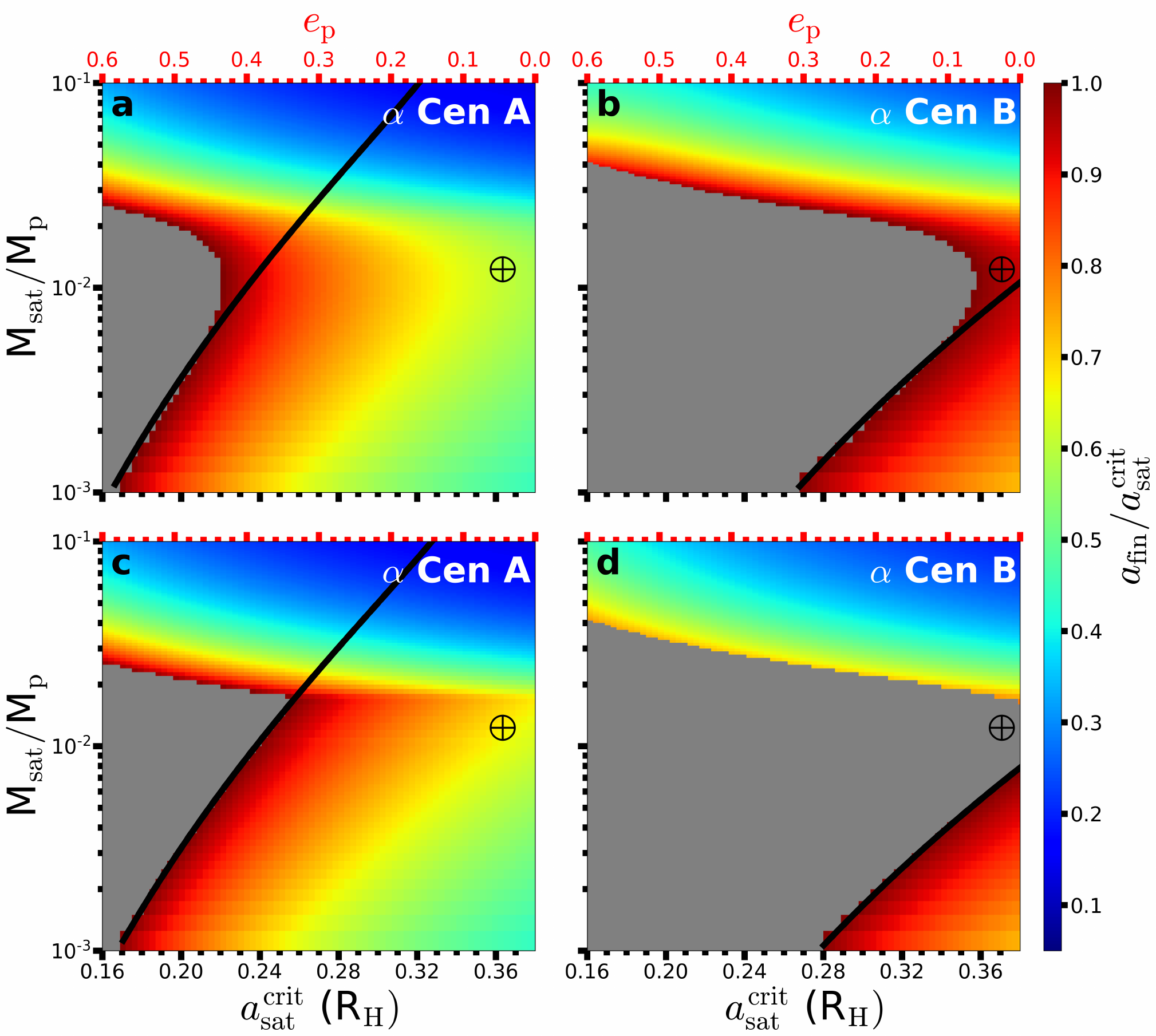}
    \caption{Tidal migration simulations over 10 Gyr using either a constant time lag (a \& b) or a constant $Q$ (c \& d) model, assuming an initially rapidly rotating (P$_{\rm rot} = 5$ hr) Earth-analog host planet with a dissipation factor $\Delta t_{\rm p} = 100 $s or $Q_{\rm p} = 33$, respectively.  The color-code denotes the final satellite semimajor axis $a_{\rm f}$ relative to the critical exomoon semimajor axis $a_{\rm sat}^{\rm crit}$ adjusted for the host planet's eccentricity (upper x-axis).  The gray region denotes parameters that allow for the satellite to escape, where the black curves mark the potential upper mass limit (see Equation \ref{eq:masslim}).  The $\oplus$ symbol denotes an Earth-Moon like mass ratio and the critical exomoon semimajor axis adjusted for the host planet's forced eccentricity (see Equation \ref{eq:eforced}).}
    \label{fig:max_moon_hz}
\end{figure}

We test the validity of this Equation \ref{eq:masslim} using an Earth-mass host planet in $\alpha$ Cen AB over 10 Gyr in a constant time lag model ($\Delta t_{\rm p} = 100$ s) following the system setup in Fig. \ref{fig:param_comp}, where the outer stability limit $a_{\rm sat}^{\rm crit}$ is truncated by the planetary eccentricity $e_{\rm p}$.  In addition, we evaluate a comparable constant $Q$ tidal model for a more direct comparison.  Figure \ref{fig:max_moon_hz} shows that Equation \ref{eq:masslim} (black curve) fits the boundary between stable (colored cells) and unstable (gray cells) for planet-satellite mass ratios $\lesssim$ 10$^{-2}$, where the assumption for only outward migration remains valid.  Above this curve, there are solutions where the infall timescale (after synchronization) is longer than 10 Gyrs.  The color-code denotes how far the satellite migrates outward over 10 Gyrs relative to the respective critical semimajor axis and the $\oplus$ symbol denotes parameters for an Earth-Moon analog at the planet's forced eccentricity.  The constant time lag models (Figs. \ref{fig:max_moon_hz}a and \ref{fig:max_moon_hz}b) allow for a smoother transition when the migration direction changes (outward to inward migration) as compared to the respective constant $Q$ models (Figs. \ref{fig:max_moon_hz}c and \ref{fig:max_moon_hz}d).  We find that Equation \ref{eq:masslim} is more applicable for planets orbiting $\alpha$ Cen B because its HZ is closer to the host star and when the planet-satellite mass ratio is $\lesssim$2-3\%.  Different assumptions on the dissipation parameter ($\Delta t_{\rm p}$ or $Q_{\rm p}$) can affect the outcome, but those also run into limits for plausibility since we are considering Earth-like values estimated from the terrestrial planets in the solar system \citep{Quarles2020b}.

\section{Observational Consequences for Planets in Binary Systems} \label{sec:observ}

\subsection{Potentially Observing Exomoons in $\alpha$ Centauri through TTVs} \label{sec:TTVs}

The presence of exomoons for single star systems can be deduced using transit timing variations (TTVs) and transit duration variations of the planet as it crosses its host star \citep{Kipping2009a,Kipping2009b}, where these measurements also depend on the planet-satellite mass ratio $m_{\rm sat}/m_{\rm p}$ and separation $a_{\rm sat}$.  These factors affect the maximum TTV {(RMS)} amplitude through the displacement of the host planet from the center of mass \citep{Sartoretti1999}, where the larger planet-satellite separations increase the TTV magnitude for a given M$_{\rm sat}/$M$_{\rm p}$.  We examine a hypothetical case of an Earth-analog near the inner edge of the conservative HZ of each star in $\alpha$ Cen (see Fig. \ref{fig:moon_hz} and how the presence of the stellar companion affects the maximum TTV amplitude for the putative transiting planet due to a prograde orbiting satellite.  The Earth-analog is assumed to be at its forced eccentricity to set the outer stability limit.

\begin{figure}
	\includegraphics[width=\columnwidth]{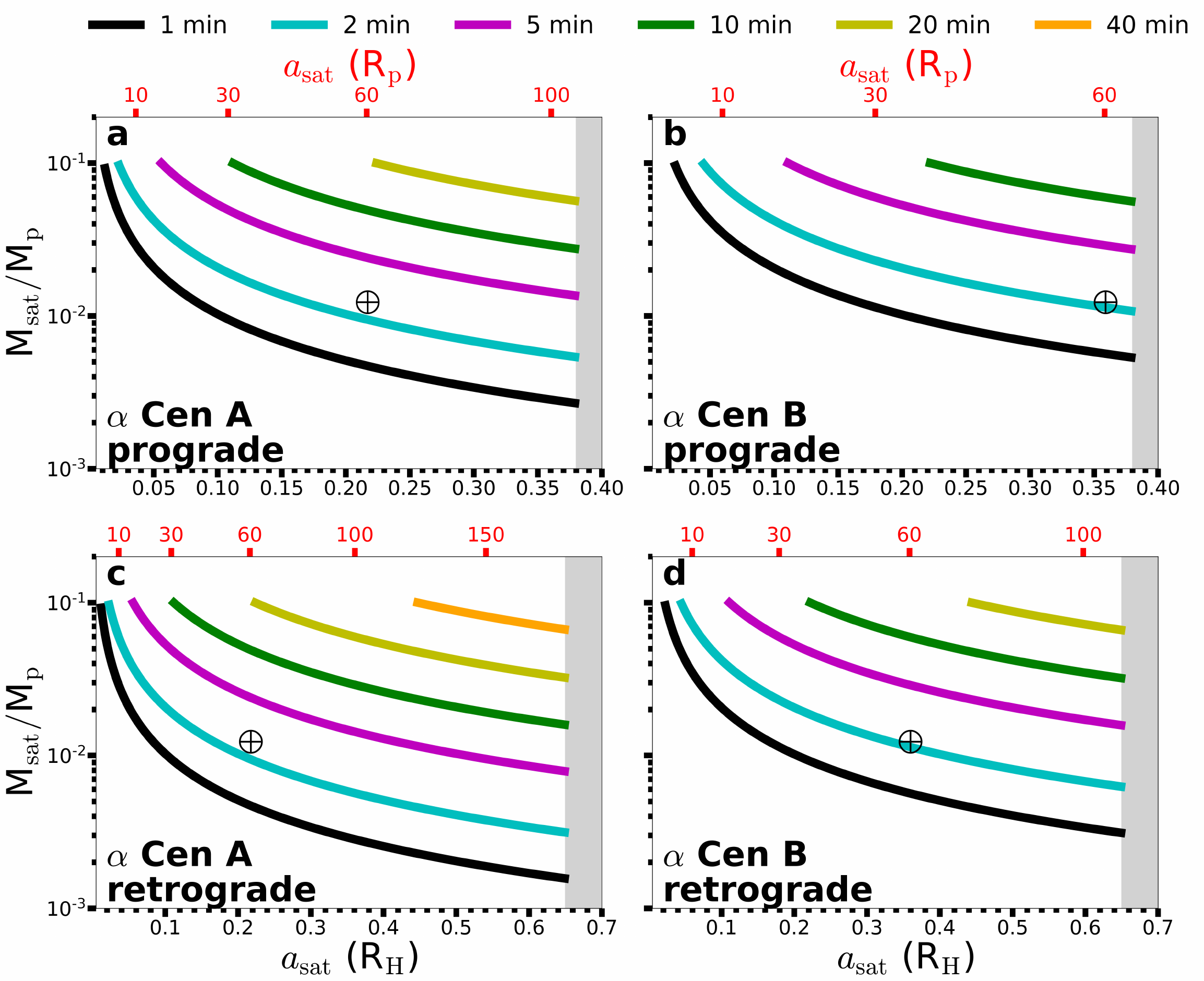}
    \caption{Potential transit timing variations (color-coded), or TTVs, for a satellite orbiting an Earth-mass planet near the inner edge of the conservative HZ of either star in $\alpha$ Cen AB.  The maximum satellite separation $a_{\rm sat}$ is truncated to account for the planet's eccentricity oscillation induced from the stellar companion, where the gray region is {not} likely to host stable exomoons.  The $\oplus$ symbol denotes a mass ratio and modern separation for the Earth-Moon system for comparison.  The upper x-axis marks the exomoon semimajor axis $a_{\rm sat}$ in units of planetary radius R$_{\rm p}$, where the bottom x-axis denotes $a_{\rm sat}$ in units of Hill radius R$_{\rm H}$. }
    \label{fig:TTV_curves}
\end{figure}

The binary companion primarily affects potential exomoons through the truncation of the outer stability limit.  Planetary tides could also limit the allowed parameters, but the tidal dissipation could be plausibly adjusted to allow for long-lived exomoons and thus cannot place a meaningful constraint for Earth-analogs in $\alpha$ Cen AB.  However, satellites of Earth-analogs in other binary systems with less luminous secondary stars can be meaningfully constrained \citep{Sasaki2012,Quarles2020b}.  Figure \ref{fig:TTV_curves} shows the parameters that can produce 1, 2, 5, 10, 20, and 40 min TTVs (color-coded) induced by prograde or retrograde satellites.   An analog of the current Earth-Moon system in terms of mass ratio and separation is denoted by the $\oplus$ symbol, which would produce a $\gtrsim$2 min TTV.  Much larger TTVs ($\lesssim$40 min) are possible for retrograde orbiting satellites because the outer stability limit is farther from the host planet.  An instrument similar to the \textit{Kepler Space Telescope} is ideal for observing circumstellar planets and measuring potential TTVs \citep{Ford2011}.  The Transiting Exoplanet Survey Satellite \citep{Ricker2016} has already detected a transiting planet within a stellar triple consisting of M-stars (LTT 1445ABC; \cite{Winters2019}), however the planet is likely already tidally locked which may limit its chances to host long-lived moons \citep{Sasaki2012}.

\subsection{Exomoons in Other Binary Systems} \label{sec:monte_carlo}
Although an Earth-analog in $\alpha$ Cen AB is a good case study, results based on the latter may not translate to other populations of binary star systems. In this section we, therefore, examine the frequency of Earth-analogs hosting moons in binary systems more generally through a Monte Carlo experiment\footnote{Processed data and python scripts underlying this article are available on\dataset[GitHub]{https://github.com/saturnaxis/exomoon-in-binaries}.  The raw data underlying this article will be shared on reasonable request to the corresponding author.}.  In our previous work \citep{Quarles2019,Quarles2020}, we used empirically derived probability density functions (PDFs) from surveys of binary stars with a Solar-Type primary \citep{Raghavan2010,Moe2017}.  The PDF for the binary period distribution is a log-normal distribution for the binary period P in days, $p_{\rm log P} \propto e^{-({\rm log P}-\xi)^2/2\sigma^2}$ where $\xi = 5.03$, $\sigma =2.28$, and $4\leq {\rm log P} \leq7$.  For the mass ratio $q = M_{\rm B}/M_{\rm A}$, we use a broken power law PDF $p_q \propto q^{\gamma_n}$, where $\gamma_1=0.3$ for $q\leq0.3$ and $\gamma_2=-0.5$ for $q>0.3$.  Following \cite{Moe2017}, we add to the PDF an excess twin fraction of 0.1 for $q\geq 0.95$ to account for the observed stellar twins.  An additional power law PDF $p_e \propto e_{\rm bin}^{0.4}$ is included to account for the binary eccentricity.  Coefficients for these PDFs are numerically determined so that the total probability is equal to unity, $\int p_{\rm x}dp = 1$.  In contrast to our previous work, we ignore the uncertainty in power law exponents as they do not substantially alter the results.

To generate random binary system parameters, we numerically determine the cumulative distribution function (CDF) for each of the PDFs described above, use a draw from a uniform distribution $\mathcal{U}(0,1)$, and then numerically determine the random variate by parallel array matching.  The planetary semimajor axis $a_{\rm p}$ is calculated at 1 Earth Flux ($S_\oplus = 1 L_\odot/(1 {\rm au})^2$) and can vary depending on which star is hosting the planet.  As the PDF for the binary mass ratio varies, we adjust the secondary star's luminosity using the common power law approximation of the mass-luminosity relation for main sequence stars.  We also assume that the planet begins at its forced eccentricity (see Equation \ref{eq:eforced}).  Our final results use the semimajor axis ratio $a_{\rm p}/a_{\rm bin}$, where the planetary semimajor axis is normalized by the binary semimajor axis $a_{\rm bin}$.  We calculate the maximum satellite mass ($M_{\rm sat}^{\rm max}$) assuming an Earth-Moon analog for tides with a 10 Gyr lifetime (see Section \ref{sec:tides}) and normalize by the planetary mass $M_{\rm p} = M_\oplus$.  We choose 50 random values for the satellite mass and semimajor axis using uniform distributions within the ranges $-3 \leq \log M_{\rm sat} \leq \log [M_{\rm sat}^{\rm max}]$ and $0.05 R_{\rm H}\leq a_{\rm sat} \leq a_{\rm sat}^{\rm crit}$.  We then use the interpolation map technique \citep{Quarles2020} to verify that the generated host planet orbit is stable and our formalism from Section \ref{sec:analytic} to determine the outer stability boundary for the exomoon  $a_{\rm sat}^{\rm crit}$.  Once the stability of the planetary and exomoon orbits are validated, the TTV {(RMS)} amplitude is calculated \citep[][see their Eqn. A27]{Kipping2009a}.

\begin{figure}
	\includegraphics[width=\columnwidth]{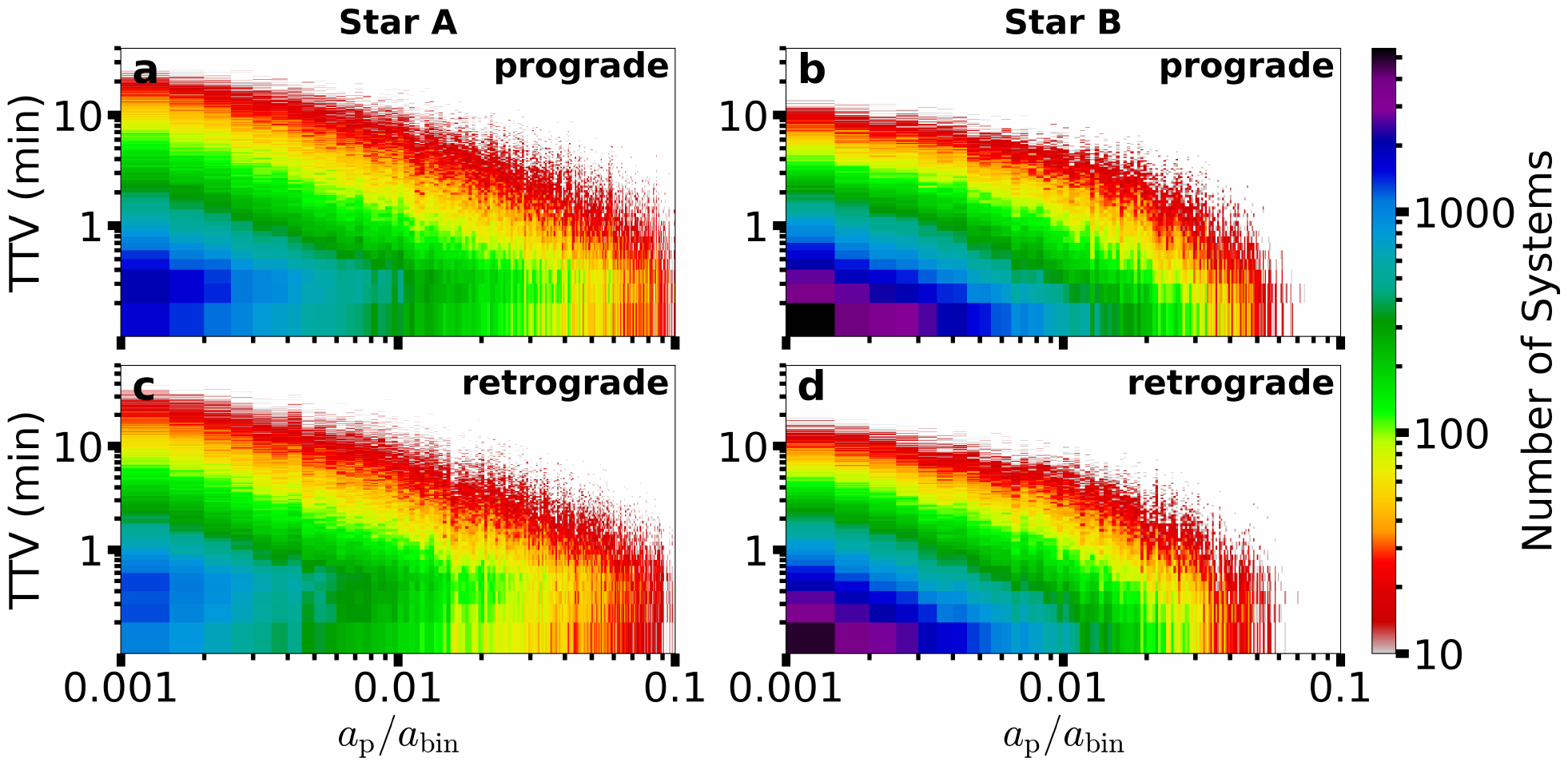}
    \caption{Monte Carlo estimation for the number of moon-hosting systems in stellar binaries with respect to the potential observables, semimajor axis ratio $a_{\rm p}/a_{\rm bin}$ and TTV {(RMS)} amplitude.  The moon-hosting planet (Earth-analog) is assumed to orbit near the inner edge of the host star's conservative HZ, while the maximum putative satellite is 0.1 M$_\oplus$.  Star A is a Solar-analog, where Star B is determined from the empirical mass ratio distribution for stellar binaries \citep{Moe2017}. The conservative HZs of binaries similar to $\alpha$ Cen AB begin at $a_{\rm p}/a_{\rm bin}$ $\sim$ 0.03--0.05, where potential exomoons can produce a 1-10 min TTV signature.   }
    \label{fig:MC_TTV}
\end{figure}

Overall, we perform 10,000 draws from the PDFs representing the binary system parameters and 50 draws of the exomoon parameters (mass and separation) assuming an Earth-analog orbiting the primary star with a prograde orbiting satellite. The corresponding results are shown in the top left panel of Figure \ref{fig:MC_TTV}. The process is repeated under the same condition, but with a retrograde orbiting satellite and prograde/retrograde satellites for an Earth-analog orbiting the secondary star to generate the results presented in the other three panels.  Figure \ref{fig:MC_TTV} illustrates the estimated number of Earth-mass planets that could host satellites for a given TTV assuming a Solar-mass primary star. The highest number of systems occurs for widely separated binary systems ($a_{\rm p}/a_{\rm bin} \sim 0.001$) and low TTV amplitudes ($\sim$0.3 min).  Higher TTV amplitudes are possible, but a large survey would likely be necessary to ensure a detection.  A retrograde orbiting exomoon around the primary star could produce a larger TTV amplitude ($\sim$30 min) due to the a larger outer stability limit, while the maximum TTV amplitude in the other cases is $\lesssim$10 min.  Counter-intuitively, planets orbiting the secondary star (B) have a slightly better chance of hosting exomoons because a distance of 1 Earth Flux is closer to the host star and less likely to be significantly perturbed by the more massive star A.  These results scale with the primary star mass, where more massive primaries (1.2 $M_\odot$) allow for larger TTV amplitudes and less massive primaries (0.8 $M_\odot$) restrict the parameter space due to more significant tidal interactions between the exomoon and host star.  The primary star of $\alpha$ Cen AB would be a good candidate for searching for TTV inducing exomoons if transiting Earth-analogs were present.  However, surveys of $\alpha$ Cen AB for planets are difficult because of pixel saturation in photometric observations \citep{Demory2015} and astrophysical noise in radial velocity observations \citep{Zhao2018}.

\section{Conclusions} \label{sec:conclusions}
Moons are ubiquitous within our solar system, where we expect similar bodies to exist in other planetary systems including those with stellar companions.  An exomoon is influenced by a strong perturber through free and forced eccentricity of its host planet.  At the minimum of the host planet's eccentricity oscillation, the planet's gravitational range of influence is truncated, which restricts the largest stable semimajor axis for the exomoon.  Additionally overlap in the MMRs between the satellite and planetary orbits allow for putative satellites to escape on relatively short timescales.  Other studies \citep{Domingos2006,Rosario-Franco2020} deduced the outer stability limit for prograde exomoons in single star systems.  We revise the stability formula for retrograde satellites as $a_{\rm sat}^{\rm crit} = 0.6684(1-1.236e_{\rm p})$ in units of Hill radii to represent the lower critical orbit.  We augment previous findings \citep{Domingos2006,Rosario-Franco2020} to include a correction using the forced eccentricity (see Equation \ref{eq:bin_stab}) determined from secular perturbation theory \citep{andrade2017secular} and thus, increase the applicability to planets in binary star systems.  

Constant time lag \citep{HUT1981} and phase lag \citep{Goldreich1966} models predict that tidal dissipation can cause outward migration to free a satellite from its host planet. We apply these models to hypothetical Earth-Moon analogs in $\alpha$ Cen AB.  The eccentricity forcing from $\alpha$ Cen A on moons hosted by an Earth-analog orbiting $\alpha$ Cen B can destabilize moons on a 10 Gyr timescale that have moderate to strong tidal dissipation ($\Delta t_{\rm p} \gtrsim 100$\,s).  In the opposite case, moons can maintain stable orbits despite strong tidal dissipation ($\Delta t_{\rm p} \sim 600$\,s).  {Although we examine coplanar orbits, \cite{Grishin2018} showed that tilted exomoon orbits can persist if the short range interactions are stronger than those from the Lidov-Kozai mechanism.}  We revise the analytic limit for planet-satellite mass fractions (see Equation \ref{eq:masslim}) in a constant phase lag (i.e., constant $Q$) and compare its validity with an equivalent constant time lag model.  The maximum planet-satellite mass fraction is most applicable to systems where the host planet's semimajor axis is more strongly influenced by the host star.  In $\alpha$ Cen AB, the conservative HZs are distant enough from the host star so that a tidal constraint is less meaningful and the formalism would be more useful for considering exomoons orbiting Proxima b \citep{Anglada-Escude2016}. 

The truncation of the Hill radius through secular eccentricity oscillations and outward tidal migration can influence potential observations of exomoons through TTVs \citep{Kipping2009a,Kipping2009b}.  The TTV {(RMS)} amplitude is largest when satellites are close to their outer stability boundaries.  These mechanisms limit the outer stability limit and can constrain the range of tidal dissipation allowed.  The maximum TTV amplitude in a system like $\alpha$ Cen AB is $\sim$40 min, where we find that an Earth-Moon analog would exhibit $\sim$2 min TTV signature.  In other binary systems, TTVs $\lesssim$1 min appear to be the most common due to trends in the underlying stellar binary population \citep{Raghavan2010,Moe2017}. {The \textit{Kepler} Mission has uncovered small TTV amplitudes ($\sim 1$ min) for many planetary systems but the significance of the TTV varies widely.  The astrophysical noise in transit photometry for circumstellar planets is expected to be similar to planets orbiting single stars as dilution from the stellar companion is less of an issue.  Current space missions like the Transiting Exoplanet Survey Satellite (TESS) focus on short period planets, which have likely already lost their moons due to the stellar tides.  Future space missions similar to \textit{Kepler} (e.g., European Space Agency's PLATO mission) that target longer period planets will be ideal in the search for exomoons.}

The search for exoplanets in binary star systems is an active field, where many targeted efforts have been applied to $\alpha$ Cen AB \citep{Endl2001,Endl2015,Bergmann2015} using the radial velocity method.  New technologies are currently in development that use direct imaging either from a coronagraph \citep{Belikov2015,Bendek2015,Thomas2015,Sirbu2017b,Belikov2017,Beichman2020}, a starshade \citep{Sirbu2017a,Bellotti2020}, or even high precision astrometry \citep{Bendek2018}.  Observations of $\alpha$ Cen AB with the Very Large Telescope (VLT) have suggested that any exoplanets there need to be $\lesssim$20 M$_\oplus$ \citep{Kasper2019}, which bodes well for the potential for terrestrial planets.  The first results of the New Earths in the $\alpha$ Centauri Region (NEAR) experiment on VLT uncovered a direct imaging signature of a roughly Neptune-sized planet orbiting $\alpha$ Centauri A \citep{Wagner2021}, but these early results still await confirmation.  Detecting exoplanets in binary star systems is a crucial step in the search for exomoons, where a wide array of methods (including TTVs) can be employed.



\acknowledgments
{The authors thank the anonymous reviewer whose comments greatly clarified and improved the quality of the manuscript.}  B.Q. thanks Maryame El-Moutamid and Rebekah Dawson for many stimulating discussions concerning tidal models.  M.R.F acknowledges support from the NRAO Gr\"{o}te Reber Fellowship and the Louis Stokes Alliance for Minority Participation Bridge Program at the University of Texas at Arlington.  This research was supported in part through research cyberinfrastructure resources and services provided by the Partnership for an Advanced Computing Environment (PACE) at the Georgia Institute of Technology. 


\software{rebound \citep{Rein2012,Rein2015}; {reboundx \citep{Tamayo2020,Baronett2021};} numpy \cite{Harris2020} }

\bibliographystyle{aasjournal}
\bibliography{refs}


\end{document}